\documentclass[final,5p,times,twocolumn,numbers]{elsarticle}

\usepackage{amssymb}
\usepackage{lipsum}
\usepackage{diagbox}
\usepackage{adjustbox}
\usepackage{tabularx}
\usepackage[table]{xcolor}
\usepackage{multirow}
\usepackage{makecell}
\usepackage[most]{tcolorbox}
\usepackage[title]{appendix}
\usepackage{url}
\usepackage{graphicx}
\usepackage{subcaption}
\usepackage{adjustbox}
\usepackage{enumitem}
\usepackage{mdframed}
\usepackage{booktabs}
\usepackage{hyperref}
\usepackage{xspace}
\usepackage{microtype}

\newcommand{\added}[1]{\textcolor{black}{#1}}

\journal{Computers \& Security}

\newcommand{\rev}[1]{\textcolor{black}{#1}}

\newcommand{\rqone}{\rev{RQ1: To what extent does SATD provide security information that complements the output of SATs?}\xspace}
\newcommand{\rqtwo}{RQ2: How do developers perceive the role of SATD when used alongside SATs for security analysis?\xspace} 

\begin{document}

\begin{frontmatter}



\title{Reading Between the Code Lines:\\ On the Use of Self-Admitted Technical Debt for Security Analysis}


\author[first,equal]{Nicolás E. Díaz Ferreyra}
\ead{nicolas.diaz-ferreyra@tuhh.de}

\author[second,equal]{Moritz Mock}
\ead{moritz.mock@student.unibz.it}

\author[first]{Max Kretschmann}
\ead{max.kretschmann@tuhh.de}

\author[second]{Barbara Russo}
\ead{barbara.russo@unibz.it}

\author[third]{Mojtaba Shahin}
\ead{mojtaba.shahin@rmit.edu.au}

\author[fourth]{Mansooreh Zahedi}
\ead{mansooreh.zahedi@unimelb.edu.au}

\author[first]{Riccardo Scandariato}
\ead{riccardo.scandariato@tuhh.de}

\fntext[equal]{Both authors contributed equally to this research as main author.}

\affiliation[first]{organization={Hamburg University of Technology},
            city={Hamburg},
            country={Germany}}

\affiliation[second]{organization={Free University of Bozen-Bolzano},
            city={Bolzano},
            country={Italy}}

\affiliation[third]{organization={RMIT University},
            city={Melbourne},
            country={Australia}}

\affiliation[fourth]{organization={The University of  Melbourne},
            city={Melbourne},
            country={Australia}}

\begin{abstract}
Static Analysis Tools (SATs) are central to security engineering activities, as they enable early identification of code weaknesses without requiring execution. However, their effectiveness is often limited by high false-positive rates and incomplete coverage of vulnerability classes. At the same time, developers frequently document security-related shortcuts and compromises as Self-Admitted Technical Debt (SATD) in software artifacts, such as code comments. While prior work has recognized SATD as a rich source of security information, it remains unclear whether -and in what ways- it is utilized during SAT-aided security analysis. 
\textbf{Objective}: \rev{This work explores whether and how the security-related information encoded in SATD provides complementary security insights to SATs.} \textbf{Method}: \rev{We followed a mixed-methods approach comprising (i) the analysis of a manually curated, SATD-annotated vulnerability dataset using three SATs and (ii) an online survey involving 72 security-aware software practitioners.} \textbf{Results}: \rev{The selected SATs flagged 114 of the 135 validated \textit{Security-related SATD instances} (SSATD), yet the overlap between SAT-derived and manually mapped Common Weakness Enumeration (CWE) identifiers was only 6.42\%, indicating that both sources often expose different kinds of security information. In particular, SSATD captured several dynamic and context-dependent weakness types that SATs commonly overlook or struggle to detect. Survey responses further indicate that practitioners rely on SSATD to contextualize SAT findings by understanding their impact, root causes, and potential fixes.} \textbf{Implications}: \rev{Our findings suggest that SSATD constitutes a valuable and cost-effective source of complementary security knowledge that can support the interpretation, prioritization, and further assessment of SAT findings.}
\end{abstract}


\end{frontmatter}

\section{INTRODUCTION} \label{sec:intro}

As code vulnerabilities continue to affect information systems, automated tools for their early and reliable detection remain a cornerstone of secure software engineering. Static Analysis Tools (SATs), such as Flawfinder~\cite{flawfinder} and Bandit~\cite{bandit}, have gained popularity among developers due to their ability to detect known vulnerabilities in source code without requiring its execution \cite{bennett2024developers,esposito2024extensive}. At their core, SATs look for syntactic and semantic patterns in code text and call sequences (e.g., hardcoded credentials, weak encryption algorithms, and improper input validations) that are indicative of security flaws \cite{croft2021empirical}. Their key value lies in surfacing information about the nature, severity, and location of these weaknesses, which is often paired with suggestions on how to fix them \cite{tahaei2021security}. Furthermore, the practical significance of SATs is evident in their widespread adoption across organizations of all sizes, driven in part by the increasing availability of open-source and lightweight alternatives~\cite{bennett2024developers,vassallo2020developers}.





\subsubsection*{\textbf{Motivation}}

Despite their popularity and value for large-scale security analysis, prior work has emphasized significant drawbacks of SATs, including high false-positive rates \cite{yedida2023find}, limited vulnerability coverage \cite{bennett2024developers,esposito2024extensive}, and persistent usability issues \cite{smith2020can,tahaei2021security,nachtigall2022large}. On the one hand, these tools trigger a large number of false alarms, which creates a major barrier for their wide-scale adoption. While several efforts have aimed to overcome this issue (e.g., through more sophisticated bug detection rules \cite{nam2019bug}), each tool usually detects only a subset of known vulnerability types \cite{bennett2024semgrep}. To mitigate this limitation, practitioners often run multiple SATs or engage in manual code inspections to identify as many vulnerabilities as possible \cite{esposito2024extensive,braz2022less,elder2022really}. However, they still struggle to interpret the results produced by these tools, which frequently lack guidance on root causes, severity, and appropriate fixes \cite{nachtigall2022large}. Consequently, many security issues are unknowingly overlooked during development and remain unresolved.

Software vulnerabilities often stem from shortcuts and compromises that developers undertake for diverse reasons, including tight delivery deadlines \cite{rindell2019sec,freire2025comprehensive}, lack of proper training \cite{coetzer2024managing}, and a narrow security mindset inside their organizations \cite{rindell2019managing}. Such insecure coding practices and shortcomings are often reported first-hand by developers as Self-Admitted Technical Debt (SATD), for instance, when explicitly referring to them inside code comments (e.g., \texttt{``TODO: Make it thread-safe''}). Recent work \cite{diaz2024satd} suggests that the information contained in SATD artifacts can support the identification and remediation of security weaknesses. That is, due to their level of detail, which, in some cases, makes it possible to map them to concrete Common Weakness Enumeration (CWE) identifiers~\cite{diaz2024satd,charoenwet2024toward}. Still, it remains unclear whether and how these insights are leveraged during SAT-aided security analysis. \rev{In particular, the extent to which security-related information encoded in SATD differs from and complements the output produced by SATs has not yet been investigated or documented in the literature, to the best of our knowledge.}

\subsubsection*{\textbf{Contribution and Research Questions}} \rev{This work presents an exploratory investigation into the use of security-related SATD (SSATD) during the identification and mitigation of security weaknesses in software projects. More specifically, we examine (i) whether the security cues contained in SSATD instances provide information that complements the output of SATs and (ii) practitioners' perspectives on how such information supports SAT-aided security analysis. In this sense, \textit{complementarity} refers to differences in the security-relevant information exposed by SAT outputs and SSATD comments, rather than to a direct comparison of vulnerability-detection performance. To investigate this phenomenon, we conducted a mixed-methods study combining the analysis of a manually curated, SATD-annotated vulnerability dataset with an online survey.} The study addresses the following Research Questions~(RQs):

\begin{itemize} [leftmargin=*]
    \item \textbf{\textit{\rqone}} To answer this RQ, we created a refined version of a pre-existing SATD-annotated vulnerability dataset \cite{MockEtAl2024Dataset} by selecting a relevant partition and enriching it with security-specific labels. Particularly, we identified SATD instances that describe security issues (i.e., code comments describing security shortcomings) and manually mapped them to concrete CWE identifiers. \rev{We then scanned the corresponding code with three SATs and compared the SAT-derived and manually mapped CWE identifiers to uncover overlaps, differences, and complementary insights.}
    \item \textbf{\textit{\rqtwo}} We addressed this RQ through an online survey targeting security-aware software developers. Overall, participants were asked whether, and in what ways, they use security-related information embedded in SSATD to complement SAT outputs. \rev{More specifically, the survey explored (i) whether SSATD information was perceived as useful for understanding and addressing selected CWEs} and (ii) what types of security insights (e.g., information on attack vectors and affected components) practitioners deem most~actionable when found in SATD artifacts.
\end{itemize}

\rev{Our findings provide exploratory evidence that security pointers available across SATD artefacts can complement the output of SATs, particularly by surfacing information that these tools may overlook. While the selected SATs flagged 84\% of the SSATD cases in our dataset, the overlap in assigned CWE identifiers was only 6.42\%, indicating that SATs and SSATD often expose different aspects of security weaknesses. Notably, the SSATD cases missed by SATs frequently describe issues of a dynamic or context-dependent nature (e.g., resource leaks or race conditions) that static analysis alone struggles to uncover. In this sense, \textbf{SSATD provides readily available security cues that can support prioritization, interpretation, and the decision to pursue deeper security assessments when warranted.}}


The remainder of this paper is organized as follows. Section~\ref{sec:background} provides the background and summarizes the related work. We explain our research methodology in Section~\ref{sec:methodology}, followed by a detailed report of the study results in Section~\ref{sec:results}. Section~\ref{sec:discussion} reflects on the findings and provides implications for research and practice. We discuss the possible threats of our study and the mitigation strategies we adopted in Section~\ref{sec:limitations}. Section~\ref{sec:conclusion} concludes the paper and discusses future research directions.

\section{BACKGROUND AND RELATED WORK} \label{sec:background}

\subsubsection*{\textbf{\textit{SAT-Supported Security Analysis}}} The use of SATs has been investigated from different angles, including their technical performance \cite{esposito2024extensive,yang2019towards,bennett2024semgrep}, usability \cite{tahaei2021security,chembakottu2025usab}, and impact on software quality and developer productivity \cite{romano2022static,elder2022really}. To a great extent, their value for security lies in their capability to identify CWEs in source code \cite{esposito2024extensive}. That is, language-agnostic software weaknesses whose causes, potential impacts, and exploitation scenarios have been systematically catalogued by MITRE. While multiple SATs can detect a shared subset of CWEs, only a few demonstrate proficiency in identifying more specific or complex ones \cite{oyetoyan2018myths}. In a recent study, Esposito et al.~\cite{esposito2024extensive} demonstrated through an extensive comparison that SATs, when combined with other vulnerability identification techniques (e.g., code inspections), provide broader CWE coverage. However, this strategy can still be impaired by the significant number of false positives they produce and the amount of warnings they report \cite{yang2019towards}. In turn, several contributions have sought to improve the precision of SATs, for instance by incorporating contextual information (e.g., argument values and types) into custom, CWE-specific detection patterns \cite{bennett2024semgrep} or heuristics \cite{RussoEtAl2022Weaksatd} based on real-world code examples. 

Despite the many efforts, SATs still struggle to detect certain CWE types like the deserialisation of untrusted data (CWE-502), input validation (CWE-20), and cross-site scripting (CWE-79) \cite{bennett2024semgrep}. Moreover, prior work has also put the usability of SATs under scrutiny, accounting for poorly designed warnings and scarce information on suitable fixes \cite{tahaei2021security,chembakottu2025usab}. Smith et al.~ \cite{smith2020can} revealed that developers often perceive SATs' feedback as overly generic or vague due to (i) the presence of inconsistent security terminology, (ii) absence of actionable fix examples, and (iii) lack of details on how vulnerabilities could be exploited. These findings have been confirmed by follow-up investigations, also stressing the importance of gathering contextual information to produce both relevant and actionable warnings \cite{nachtigall2022large,charoenwet2024toward}. Particularly, emphasis has been placed on leveraging insights from code review comments to improve SATs's configurations (e.g., disabling irrelevant checks) and reduce false alarms \cite{zampetti2022using}. Furthermore, a recent study by Charoenwet et al.~\cite{charoenwet2024toward} on two large open-source projects found that such comments can provide actionable information across multiple weakness categories, including authentication errors, privilege issues, and behavioural flaws. Yet, it remains unclear to what extent this knowledge could help overcome some of SATs' technical limitations and address recurring usability challenges documented in the current literature.

\subsubsection*{\textbf{\textit{SATD for Security Assessment}}} Over the last decade, the study of SATD has attracted significant attention from researchers concerned with the quality and long-term maintainability of software systems. Thanks to their inherent natural language form, SATD instances have served as proxies for detecting issues stemming from unfinished requirements, design flaws, and code defects \cite{obrien202223}. Overall, SATD research spans across different domains, including Machine Learning (ML) \cite{pepe2024taxonomy}, scientific software \cite{hassan2025characterising}, and architectural sustainability \cite{sutoyo2025tracing,ojameruaye2016sustainability}. Early investigations have elaborated on methods and tools for the identification of SATD instances, ranging from pattern matching techniques to transformer-based ML algorithms \cite{gu2024satd}. For instance, Potdar and Shihab~\cite{PotdarShihab2014SATD} introduced an initial set of 62 keywords commonly found across SATD comments, which was later extended with task annotation tags (e.g., \texttt{``XXX''}, \texttt{``TODO''}, and \texttt{``FIXME''}) by Guo et al.~\cite{GuoEtAl2021MAT}. Deep Learning (DL) models (e.g., BERT) have also been adopted to overcome the limitations of pattern-based methods (e.g., low recall) \cite{gu2024satd} and to expand the identification of SATD beyond code comments (e.g., to pull requests and commit messages) \cite{li2023automatic}. Furthermore, recent work has explored the use of Large Language Models (LLMs) like GPT-4 to improve the classification of SATD observations \cite{sheikhaei2024empirical,li2025impact} and make SATD identification more context-aware \cite{yonekura2025context}.

The management of Security Debt (SD) in software projects has also gained increasing interest from both researchers and practitioners in the last years \cite{kruke2024defining}. Similar to other, more established types of Technical Debt (TD), SD encompasses security issues that origin from (simple) coding mistakes and suboptimal solutions that tend to accumulate over time. Prior work on SD spans across its use for risk estimations \cite{rindell2019sec}, vulnerability prediction \cite{siavvas2022technical}, along with its role in the prioritization and remediation of security fixes \cite{izurieta2018td}. Still, research at the intersection of SD and SATD remains limited and relatively recent. For example, \cite{RussoEtAl2022Weaksatd} observed that more than 55\% of SATD comments available in the Chromium project\footnote{\url{https://www.chromium.org/Home/}} correlate to insecure C-code implementations, whereas \cite{diaz2024satd} showed that SATD artifacts referencing security issues often provide enough detail to spot CWEs in Open Source Software (OSS) projects. The latter study also found, through an online survey, that developers frequently report security weaknesses as SATD to support the timely identification and fixing of vulnerabilities. Nevertheless, to the best of our knowledge, prior investigations have not yet elaborated on the interplay between security-relevant SATD and the output produced by SATs. Particularly, on whether, and how, the security references available in SATD artifacts can contribute (or play a role) during SAT-supported security analyses.

\section{METHODOLOGY}\label{sec:methodology}

To answer the RQs formulated in Section~\ref{sec:intro}, we followed a mixed-methods approach encompassing (i) the analysis of a SATD-annotated vulnerability dataset and (ii) an online survey with security practitioners. The following sections describe the steps we followed at each stage along with the instruments employed during the different data acquisition and processing activities.

\begin{figure}
    \centering
    \includegraphics[width=\linewidth]{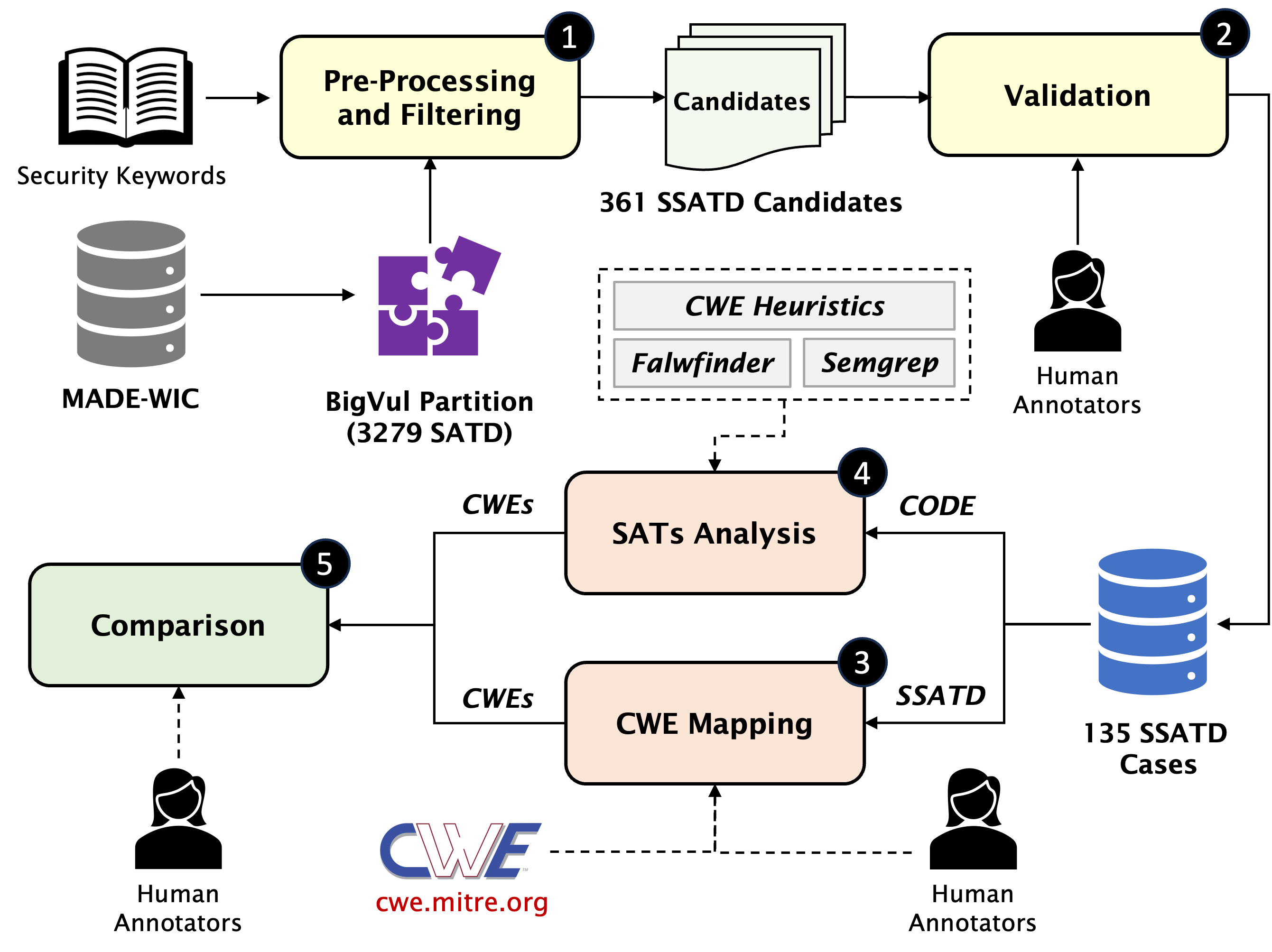}
    \caption{Methodology applied for the Dataset Study.}
    \label{fig:methodology}
\end{figure}

\subsection{Dataset Study}

This first study explored the security-related information encoded in SATD artefacts and its relationship to the outputs of different SATs. \rev{To this end, we conducted a keyword-based search on a pre-existing SATD-annotated dataset to identify candidate SSATD instances, which were subsequently validated manually and mapped to CWE identifiers. We then analyzed the corresponding source code using three SATs and compared the SAT-derived and manually mapped CWE identifiers.}

\subsubsection*{\textbf{Reference Dataset}} We leveraged the MADE-WIC dataset by Mock et al.~\cite{MockEtAl2024Dataset}, which comprises 859,774 functions spanning across multiple C and C++ open source projects. MADE-WIC is structured into three subsets or partitions, from which two correspond to pre-existing vulnerability datasets, namely Devign \cite{ZhouEtAl2019Devign} and Big-Vul \cite{FanEtAl2020Big-Vul}. One distinctive feature of MADE-WIC is that, unlike Devign and Big-Vul, it includes the leading comments of each function (i.e., those directly preceding a function's header) along with their corresponding SATD annotations (i.e., according to the patterns and tags introduced by Potdar and Shihab~\cite{PotdarShihab2014SATD} and Guo et al.~\cite{GuoEtAl2021MAT}). We chose the Big-Vul partition for this study as it covers the highest number of open-source projects (some of them included in the other two), while offering a good ratio of SATD-annotated functions (\textbf{3,279~cases}).

\subsubsection{\textbf{STEP 1: Pre-Processing and Filtering}} Fig.~\ref{fig:methodology} illustrates the steps followed to curate a dataset of Security Self-Admitted Technical Debt (\textbf{SSATD}), namely SATD instances that describe security issues in their corresponding code sections. First, using a pre-existing list of security terms by Croft et al.~\cite{CroftEtAl2022SecI}, we conducted a keyword-based search among the SATD cases in MADE-WIC's Big-Vul partition. This list is an updated version of the one proposed by Le et al.~\cite{le2020puminer} and contains 288 security keywords (e.g., `leak' and `thread-safe'), making it one of the most extensive lists documented in the literature. \rev{The keyword search was used only to identify \textit{candidate} SSATD instances for subsequent manual validation and did not determine their final classification. Furthermore, although keyword-based filtering may overlook cases that do not contain explicit security terminology and may also produce false positives, it provides a transparent and reproducible mechanism for narrowing the initial set of 3,279 SATD instances.} We applied the search after preprocessing the leading comments of each SATD-positive instance by removing punctuation characters, obtaining a total of \textbf{361 SSATD candidates}.

\subsubsection{\textbf{STEP 2: Validation}} Two authors with prior experience in software security independently validated the 361 SSATD candidates by checking for false positives. Following a shared validation criterion, they assessed whether the information disclosed in each leading comment could indicate the presence of a security flaw. Each candidate was labelled as either a \textit{true positive} (TP) or a \textit{false positive} (FP), and agreement between the two classifications was measured using Cohen's Kappa. \rev{The resulting coefficient of 0.34 indicates `fair' agreement \cite{landis1977measurement}, highlighting the challenges of consistently classifying SSATD.} The authors subsequently reviewed all disagreements individually and resolved them through discussion until consensus was reached. This process resulted in a final dataset of \textbf{135 validated SSATD instances}.

\subsubsection{\textbf{STEP 3: CWE Mapping}} \label{sec:manual_cwe} After assessing and validating the SSATD candidates, the same authors mapped each of the 135 true positive cases to Common Weakness Enumeration Identifiers (CWE-IDs) in a follow-up session. Supported by MITRE, the CWE initiative consists of a public, community-developed catalogue with information about more than 900 types of software and hardware weaknesses. The mapping was conducted following an exploratory, \rev{manually conducted procedure}: one author with security expertise performed the initial mapping of the 135 validated SSATD instances to CWE identifiers by browsing the CWE database\footnote{\url{https://cwe.mitre.org/}} using the keywords found in each SSATD item as guidance. A second author, also a security expert, \rev{reviewed} all mappings. Disagreements occurred in only four cases and were resolved through joint discussion until consensus was reached. We prioritized mappings to ALLOWED CWEs (i.e., current and active weakness categories) and resorted to DISCOURAGED CWEs (i.e., still valid but no longer recommended for use) only when no clear match was available. Mapping to PROHIBITED CWEs was explicitly avoided.

\subsubsection{\textbf{STEP 4: SATs Analysis}} \label{sec:sat_cwe} We ran multiple SATs on the source code associated with each SSATD instance to obtain tool-generated CWE mappings for comparison with the manually mapped SSATD information. We selected tools that (i) supported C/C++ analysis, (ii) provided built-in mappings to MITRE's CWE repository, and (iii) were available as open-source software. \rev{Because the reference dataset \cite{MockEtAl2024Dataset} consists of extracted functions that are not independently executable, we additionally required tools capable of analysing incomplete or non-executable code. These criteria enabled a standardized comparison based on CWE identifiers and supported the reproducibility of the study.} The following SATs were selected and applied:

\begin{itemize}[leftmargin=*]
    \item \textbf{Semgrep:} A lightweight, open-source SAT designed to detect security vulnerabilities, software bugs, and code quality issues across more than 30 programming languages. It includes 53 built-in rules for C/C++, which can be extended with custom or community-provided ones. Semgrep operates through pattern-based code matching (i.e., via Abstract Syntax Trees), making it fast and suitable for scanning incomplete or non-compilable code \cite{ImprotaEtAl2025semgrep}.
    \item \textbf{Flawfinder:} This SAT employs an internal database of security-relevant functions commonly associated with vulnerabilities (e.g., \texttt{strcpy}, \texttt{gets}, \texttt{sprintf}) and their corresponding CWE identifiers. The tool operates by extracting all function calls from the input code (i.e., whether spanning multiple lines, individual functions, or entire files) and cross-referencing them with entries in its database. Currently, Flawfinder’s database includes 222 security-relevant functions, each linked to one or more CWE identifiers \cite{flawfinder}.
    \item \textbf{CWE Heuristics:} Proposed by Russo et al.~\cite{RussoEtAl2022Weaksatd}, this method detects weak code through an example-based search strategy, extending beyond simple signature matching approaches such as those used in Flawfinder. Particularly, functions are identified through specific code examples (i.e., problematic code patterns or constellations) available in MITRE's CWE database (e.g., CWE-416\footnote{\href{https://cwe.mitre.org/data/definitions/416}{cwe.mitre.org/data/definitions/416}} or CWE-484\footnote{\href{https://cwe.mitre.org/data/definitions/484}{cwe.mitre.org/data/definitions/484}}), allowing for a wider range of CWE~detection.
\end{itemize}


\rev{We ran the selected SATs using their default configurations to represent a common out-of-the-box usage scenario and facilitate replication of the reported findings among researchers and practitioners \cite{bennett2024developers,vassallo2020developers}. Consequently, the results characterize the outputs produced by the selected tools under these configurations and should not be interpreted as estimates of their maximum detection capabilities.}

\subsubsection{\textbf{STEP 5: Comparison}}\label{met:comparison} We compared the CWEs obtained through the manual SSATD mapping with those produced by the three SATs, namely Semgrep, Flawfinder, and CWE Heuristics. \rev{Because SATs inspect source code whereas SSATD mappings are derived from developer-written comments, our analysis does not constitute a one-to-one comparison of vulnerability-detection performance. Rather, it examines the extent to which the two approaches expose overlapping or distinct security-relevant information.}

Since a given weakness may be associated with different CWE-IDs by the SSATD annotators (STEP 3) and the SATs (STEP 4), we extended the analysis beyond exact identifier matches to consider relationships between the corresponding CWE descriptions. We first generated all possible pairs between the CWEs mapped from SSATD and those reported by each SAT for the same instance, resulting in a one-to-many relationship. Two authors independently assessed each pair to determine whether the corresponding CWEs were related, based on the descriptions and code examples available in MITRE's CWE database. Following the initial independent assessment, all disagreements were discussed and resolved through consensus. Inter-rater reliability before consensus was measured using Cohen's Kappa and yielded a coefficient of 0.672, indicating `substantial' agreement \cite{landis1977measurement}.

\subsection{Survey Study} \label{sec:survey_study}
To gather further insight's into the interplay between SSATD instances and the output produced by SATs, we conducted an online survey with security practitioners. It consisted of 7 blocks of closed-ended questions (plus an open ended-one) concerning the use of SATs along with the information available in SSATD instances to support the identification and fixing of certain security weaknesses (i.e., specific CWE-IDs).

\begin{figure}
    \centering\small
    \begin{mdframed}[backgroundcolor=gray!10]
    \textbf{QUESTION}: Please indicate how often you IDENTIFY the following SECURITY WEAKNESS with the help of Static Analysis Tools (SATs):\vspace{1ex}
    
    \textbf{Race Condition (CWE-362)}: \textit{This issue happens when two or more processes try to use the same resource at the same time without proper coordination. It can cause unpredictable behaviour, data corruption, or security vulnerabilities.}\vspace{1ex}
    
    \textbf{ANSWER OPTIONS}: \\\texttt{NEVER | RARELY | SOMETIMES | OFTEN | ALWAYS}
    \end{mdframed}\vspace{-2ex}
    \caption{Description used in the survey for CWE-362.}
    \label{fig:vignete}
\end{figure}

\subsubsection{Structure}\label{sec:surv_str} Prior work by Díaz Ferreyra et al.~\cite{diaz2024satd} revealed that references to CWE-362 (\textit{`Race Condition'}), CWE-119 (\textit{`Improper Restriction of Operations within the Bounds of a Memory Buffer'}), CWE-402 (\textit{`Resource Leak'}), CWE-20 (\textit{`Improper Input Validation'}), and CWE-404 (\textit{`Improper Resource Shutdown or Release'}) were among the most frequent ones across a dataset of 201 SSATD instances. Building upon these findings, we designed a survey divided in two parts to capture developer's perspective about the role of both SATs and SSATD in the identification and resolution of these five particular~CWEs. 
\begin{itemize} [leftmargin=*]
    \item \textbf{PART 1}: We asked the participants how frequently SATs help them identify these CWEs, and the extent to which SAT outputs contribute to understand and remediate their associated weaknesses. Each question was formulated in a Likert style and guided by a short description of each CWE (as shown in Fig.~\ref{fig:vignete}). Participants were instructed to answer based on the SAT they were most familiar with.
    \item \textbf{PART 2}: We asked whether security references available in SATD artefacts (e.g., code comments) provide valuable insights to these activities. As in PART 1, each question was formulated in a Likert style and guided by a short description of each CWE, but this time framed towards the role of SSATD (e.g., ``The security pointers contained inside software artefacts help me UNDERSTAND the root cause of this weakness''). Finally, we investigated the extent to which SSATD instances help developers, in general, to better understand the characteristics of the CWEs detected by SATs, namely their (i) \textit{type}, (ii) \textit{impact}, (iii) \textit{affected software components}, and (iv) potential \textit{attack vectors}. For this purpose, we included one closed-ended question for each characteristic, along with an open-ended item to capture additional insights.
\end{itemize}

A subset of the main survey questions can be found in \autoref{appendix}. The complete survey instrument is available in the paper's \hyperref[sec:replication]{\textbf{Replication Package}}.

\subsubsection{Population and Recruitment}\label{sec:population}

Participants were recruited via Prolific\footnote{\url{https://www.prolific.com/}}, a crowdsourcing platform frequently used for conducting empirical studies in software engineering \cite{tahaei2022recruiting,kaur2022recruit}. We targeted individuals who, in their Prolific profiles, reported having (i) knowledge of software development techniques and (ii) computer programming skills. These qualifications were validated through a \textbf{short technical questionnaire}, similar to the one proposed by Krause et al.~\cite{krause2023pushed}, consisting of \textbf{five closed-ended questions}.
\begin{enumerate}[label=(\roman*), noitemsep, topsep=0pt]
    \item The \textbf{first three questions} assessed basic programming comprehension by asking for the outputs and parameters of simple pseudocode snippets (e.g., merging two lists). These snippets were displayed as images to make it difficult for participants to generate responses using Generative Artificial Intelligence (GenAI), considering that models such as ChatGPT limit the number of free image-based queries per day. \added{Participants who failed to answer all three questions correctly were excluded during data analysis.}
    \item \added{The \textbf{final two questions} measured participants' knowledge of software security and their familiarity with SATs. In the first one, they were given a list of security-relevant features (e.g., encryption) and asked if they had ever incorporated them into their code. Similarly, they were asked about the use of different code debugging techniques, including code scanning tools. Participants who never implemented a security feature or ran a code analysis tool were disqualified.}
\end{enumerate} 
This screening questionnaire took around 3 minutes, for which we rewarded participants with 0.3 GBP each. Only those who passed it were invited to take the \textit{main survey} (Section~\ref{sec:surv_str}) and received an additional compensation of 1.15 GBP. As a general rule, participants had to be at least 18 years old to join both parts of the study (i.e., screening questionnaire and main survey), have taken part in at least 10 other studies in Prolific, and have a minimum approval rate of 98\%. Two attention questions were also included in the main survey to identify and discard answers from unengaged participants.

\textbf{\textit{\added{Sample size.}}} After running a pilot study with 10 participants, we observed that around 30\% passed the technical questionnaire. Based on this ratio and our study budget, we targeted a sample of size N=70. Hence, we repeated the process with 200 additional participants, from which 81 were deemed suitable for undertaking the main survey (i.e., after validating their technical knowledge and level of security expertise). Of these 81 participants, 8 did not accept our invitation to take the main survey, resulting in 73 main survey submissions. After checking for completeness and attention questions, a total of \textbf{72 answers} remained valid for analysis.

\subsubsection{Ethical Considerations}

We received approval from the German Association for Experimental Economic Research to conduct this study and followed the guidelines prescribed in the Declaration of Helsinki. We informed participants about the study procedure along with details concerning the privacy of their personal data before joining the experiment. We also asked for their informed consent and allowed them to withdraw at any time without their answers being recorded.

\section{RESULTS} \label{sec:results}

We performed a series of analyses on the collected SSATD instances and the survey responses. The key findings, along with the main characteristics of the resulting datasets, are presented in the following subsections.



\begin{table}
    \centering\small
    \caption{Number of flagged cases and CWE types per method.}
    \begin{tabularx}{\columnwidth}{>{\centering\arraybackslash}X *{2}{>{\centering\arraybackslash}X}}
        \toprule
        \textbf{Method} & \textbf{Flagged Cases} & \textbf{CWE-IDs} \\
        \midrule
        CWE Heuristics  & 113 & 16 \\
        Flawfinder      & 24  & 8  \\ 
        \rowcolor{gray!25}\textbf{All SATs combined}        & \textbf{114} & \textbf{24} \\ 
        \rowcolor{gray!25}\textbf{SSATD}           & \textbf{135} & \textbf{33} \\
        \bottomrule
    \end{tabularx}
    \label{tab:total_flagged}
\end{table}

\begin{table}
    \centering\small
    \caption{Method-exclusive CWE types.}
    \begin{tabularx}{\columnwidth}{@{}lX@{}}\toprule
         \textbf{Method} & \textbf{CWE-IDs}  \\ \hline
         \textbf{SATs} & \mbox{CWE-126}, \mbox{CWE-78}, \mbox{CWE-190}, \mbox{CWE-134}, \mbox{CWE-807}, \mbox{CWE-327}, \mbox{CWE-783}, \mbox{CWE-195}, \mbox{CWE-196}, \mbox{CWE-676}, \mbox{CWE-467}, \mbox{CWE-244}, \mbox{CWE-415}, \mbox{CWE-483}, \mbox{CWE-14}, \mbox{CWE-690}, \mbox{CWE-478}, \mbox{CWE-135}, \mbox{CWE-587}, \mbox{CWE-806} \\\midrule
         \textbf{SSATD} & \mbox{CWE-758}, \mbox{CWE-833}, \mbox{CWE-118}, \mbox{CWE-754}, \mbox{CWE-805}, \mbox{CWE-392}, \mbox{CWE-664}, \mbox{CWE-665}, \mbox{CWE-799}, \mbox{CWE-230}, \mbox{CWE-284}, \mbox{CWE-362}, \mbox{CWE-662}, \mbox{CWE-602}, \mbox{CWE-200}, \mbox{CWE-705}, \mbox{CWE-404}, \mbox{CWE-358}, \mbox{CWE-79}, \mbox{CWE-153}, \mbox{CWE-601}, \mbox{CWE-354}, \mbox{CWE-240}, \mbox{CWE-400}, \mbox{CWE-325}, \mbox{CWE-667}, \mbox{CWE-402}, \mbox{CWE-131}, \mbox{CWE-1188} \\\bottomrule
    \end{tabularx}
    \label{tab:method_cwes}
\end{table}

\begin{table*}
    \centering\small
    \caption{CWE types of the unique SSATD-flagged cases.}
    \begin{tabularx}{\linewidth}{p{5.5cm} X}\toprule
         \textbf{CWE-ID} & \textbf{SSATD Example}  \\ \hline
         \rowcolor{gray!25}\textbf{CWE-665$^{*}$}: Improper Initialization  &  \texttt{``TODO: Use the real values once the openssl constants are used'' }\\
         \textbf{CWE-362$^{*}$}: Race Condition &  \texttt{``TODO: ... We might need to open the file on startup...,otherwise there would be a race condition.''} \\
         \rowcolor{gray!25}\textbf{CWE-758$^{*}$}: Reliance on Undefined, Unspecified, or Implementation-Defined Behavior & \texttt{``XXX we don't know yet if the IANA will preclude overlap of 1-byte and 2-byte spaces. If not, we need to offset tag after this step.''}\\
         \textbf{CWE-358$^{*}$}: Improperly Implemented Security Check for Standard &  \texttt{``TODO: ...Check if a user decision has been made to allow or deny SSL certificates with errors on this site.''}\\
         \rowcolor{gray!25}\textbf{CWE-354$^{*}$}: Improper Validation of Integrity Check Value & \texttt{``TODO: Should check and handle checksum.''}\\
         \textbf{CWE-402$^{*}$}: Resource Leak & \texttt{``TODO: lldds need to unconditionally forget about aborted ata tasks, otherwise we (likely) leak the sas task here''}\\
         \rowcolor{gray!25}\textbf{CWE-754}: Improper Check for Unusual or Exceptional Conditions & \texttt{``XXX we could spend more on the wire to get more robust failure detection, arguably worth it to avoid data corruption''}\\
         \textbf{CWE-20}: Improper Input Validation & \texttt{``TODO: Validate the QName''}\\
         \rowcolor{gray!25}\textbf{CWE-401}: Missing Release of Memory after Effective Lifetime & \texttt{``XXX - page\_mkwrite gets called every time the page is dirtied, even if it was already dirty, so for space accounting reasons we need to clear any delalloc bits''}\\
         \bottomrule
         \multicolumn{2}{l}{(*) CWEs identified exclusively through manual SSATD mapping.}  \\
    \end{tabularx}
    \label{tab:ssatd_cwes}
\end{table*}










\subsection{Detection of security weaknesses (RQ1)} \label{sec:res_1}

As shown in Table~\ref{tab:total_flagged}, the evaluated SATs (i.e., Flawfinder, Semgrep, and the CWE Heuristics) detected, altogether, 114 of the 135 SSATD cases in our dataset (i.e., flagged the respective code section as insecure). Surprisingly, while Flawfinder and the CWE Heuristics managed to identify 24 and 113 cases of security weaknesses, respectively, Semgrep found none. Furthermore, only one instance was simultaneously detected through both the CWE Heuristics and Flawfinder.  Altogether, the 114 SAT-flagged dataset instances covered 24 distinct CWE types (i.e., 16 detected by the CWE Heuristics and 8 by Flawfinder), with no overlap between them. On the other hand, from the 135 SSATD cases, 115 were successfully mapped manually to CWE-IDs accounting for 33 different types of security weaknesses. Overall, we identified 53 different CWE-IDs across all methods (i.e., from STEP 3 and STEP 4) from which 20 were exclusively identified by a SAT and 29 exclusively through the manual SSATD mapping (Table~\ref{tab:method_cwes}). Only 4 CWE types are common to both approaches, namely \textit{CWE-120: `Classif Buffer Overflow'}, \textit{CWE-20: `Improper Input Validation'}, CWE-416: \textit{`Use After Free'}, and \textit{CWE-401: `Missing Release of Memory after Effective Lifetime'}.

\begin{tcolorbox}[colback=gray!20, boxsep=1pt]
\textbf{Findings 1}: The evaluated SATs managed to flag, altogether, 84\% of the SSATD instances in our dataset and assigned them to 24 different CWE-IDs. The manual mapping of SSATD comments spanned across 33 different CWE types. Only 4 CWE-IDs are common to both approaches.
\end{tcolorbox}

As mentioned in Section~\ref{met:comparison}, to better understand the correspondence between SSATD comments and SAT-reported weaknesses, we further examined all cases where both the manual annotations (Section~\ref{sec:manual_cwe}) and the SATs (Section~\ref{sec:sat_cwe}) assigned a CWE identifier. This resulted in the analysis of 147 unique pairs of CWE-IDs from which only 6.42\% of them were assessed as ``closely related''. One of such cases concerns \textit{CWE-244: `Heap Inspection'} and \textit{CWE-664: `Improper Control of a Resource Through its Lifetime'}, respectively identified by a manual SSATD mapping and a SAT. Although the former addresses a more specific issue in which a heap is not properly cleared after used, they both concern the management of resources across their lifetime. Conversely, pairs like \textit{CWE-783: Operator Precedence Logic Error} and \textit{CWE-362: Race Condition} were judged as ``unrelated'', accounting for 85.71\% of the cases. In this example, whereas the former refers to logical mistakes in operator ordering, the latter describes concurrency issues involving shared resources. While one might argue that precedence errors could indirectly affect concurrent behaviour, both raters agreed that such an overlap is coincidental rather than systematic. For the remaining 8.16\% of the pairs, no consensus could be reached.

\begin{tcolorbox}[colback=gray!20, boxsep=1pt]
\textbf{Findings 2}: Only 6.42\% of the CWE-IDs assigned by SATs were  related to those obtained through manual SSATD mapping. Moreover, 21 out of 135 SSATD cases were identified exclusively through SSATD comment analysis, spanning 9 types of security weaknesses (6 of which were detected solely through manual SSATD examination).
\end{tcolorbox}
Of the 135 instances in our dataset, 21 were identified exclusively through SSATD (i.e., no other SAT flagged them). As shown in Table~\ref{tab:ssatd_cwes}, these cases spanned nine different CWE-IDs (i.e., CWE-665, CWE-362, CWE-754, CWE-758, CWE-20, CWE-358, CWE-354, CWE-401, and CWE-402), six of which correspond to types of security weaknesses only identified through SSATD mapping (CWE-665, CWE-362, CWE-758, CWE-358, CWE-354, and CWE-402). These weaknesses primarily capture issues of dynamic nature that are inherently difficult (if not impossible) for SATs to identify. For instance, CWE-358: \textit{`Improperly Implemented Security Check for Standard'} or \textit{CWE-402: `Resource Leak'} often manifest through runtime conditions or context-dependent logic that SATs cannot reliably infer. These cases illustrate the added value of SSATD: comments may convey highly specific information about potential vulnerabilities that remain beyond the reach of current, static-based analysis techniques.

\subsection{Developers' practices and standpoints (RQ2)}

As mentioned in Section~\ref{sec:population}, we received 72 valid responses, of which 54 corresponded to male participants, 16 to female, and 2 to non-binary. Approximately 52.8\% reported having \textit{above-average} or \textit{very high} security skills, while 44.4\% rated their expertise as \textit{average} and only 2.8\% as \textit{below average} (Table~\ref{tab:demographics}). When asked about the SAT they were most familiar with, CodeQL, SonarQube, CodeChecker, and SpotBugs were the most frequently mentioned, accounting for 27.8\%, 26.4\%, 15.3\%, and 9.7\% of responses, respectively. Other tools, such as Semgrep, Cppcheck, and Bandit, were also mentioned but with lower frequency (i.e., each below 5\%).

\begin{table}[ht]
\caption{Survey self-reported demographic data.}
\label{tab:demographics}
\centering\small
\begin{tabularx}{\linewidth}{lXcc}
\toprule
\textbf{Demographic} & \textbf{Ranges} & \textbf{Freq.} &\textbf{\%} \\
\midrule
\multirow{3}{*}{Gender}
& Male          & 54    & 75\%   \\
& Female        & 16    & 22.2\%   \\
& Non-Binary    & 2     & 2.8\%   \\          
\midrule
\multirow{4}{*}{\makecell[l]{Educational\\level}}
& High School or Less           & 1    & 1.4\%   \\
& Some College                  & 6    & 8.3\%   \\
& Undergraduate (BSc,~BA)       & 40    & 55.6\%   \\
& Graduate (MSc, PhD)           & 25    & 34.7\%   \\
\midrule
\multirow{4}{*}{\makecell[l]{Security\\Skills}}
& Bellow average           & 2    & 2.8\%   \\
& Average                  & 32    & 44.4\%   \\
& Above average       & 29    & 40.3\%   \\
& Very high           & 9    & 12.5\%   \\
\bottomrule
\end{tabularx}
\vspace{-2ex}
\end{table}

\subsubsection{SAT-Driven security analysis} 
\begin{figure}
    \centering
    \includegraphics[width=\linewidth]{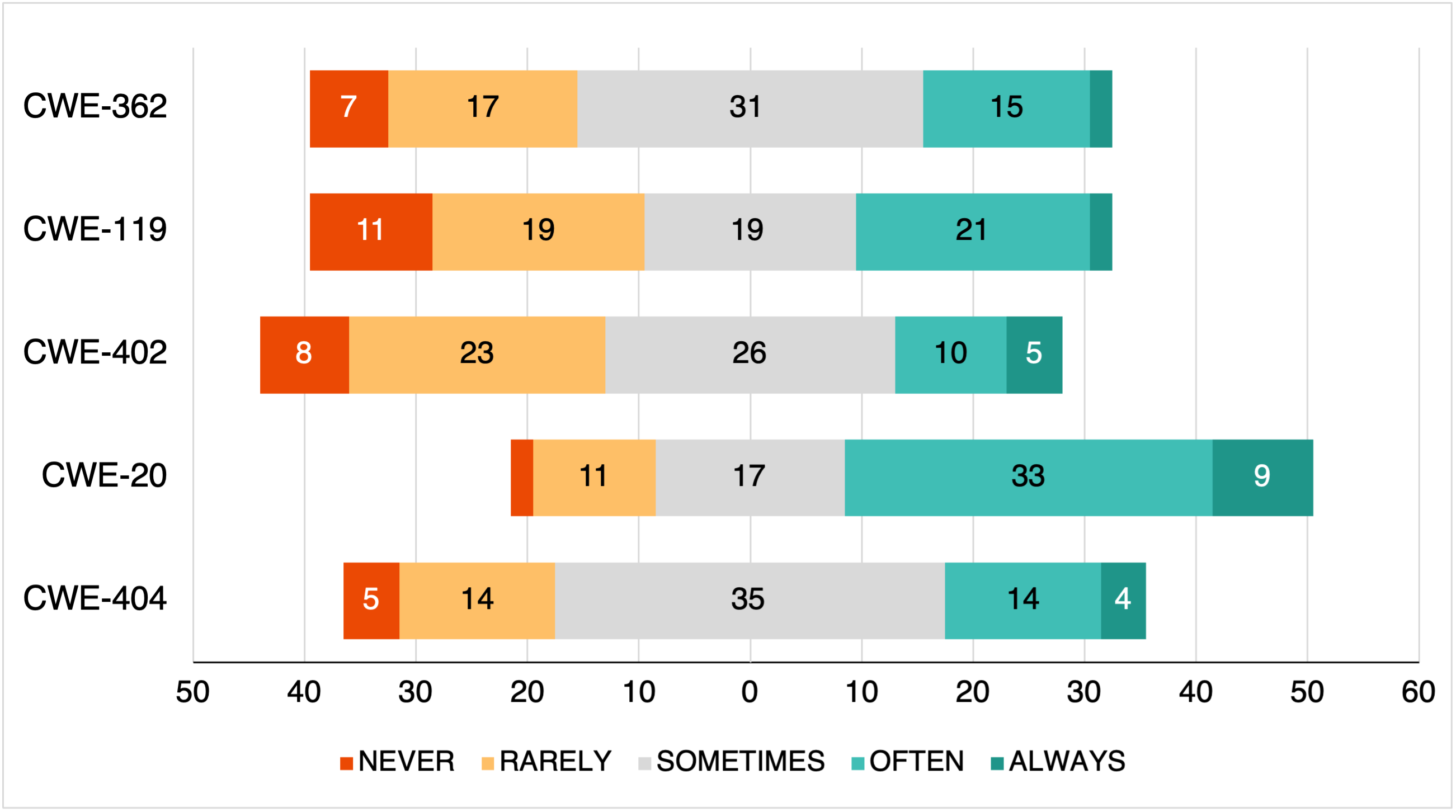}
    \caption{SATs for CWE identification.}
    \label{fig:sat_identify}
\end{figure}

From Fig.~\ref{fig:sat_identify}, we can observe that between 50\% and 68\% of participants reported using SATs \textit{sometimes} or \textit{often} to identify each of the five given CWEs. In particular, 58.3\% of respondents considered SATs to be \textit{often} or \textit{always} useful for detecting cases of CWE-20 (\textit{`Improper Input Validation`'}). Conversely, 41.6\% indicated that SATs \textit{rarely} or \textit{never} help in identifying CWE-119 (\textit{`Improper Restriction of Operations within the Bounds of a Memory Buffer'}), and 43\% expressed the same view for CWE-402 (\textit{`Resource Leak'}).

Participants were also asked to assess whether SATs provide relevant information to UNDERSTAND and FIX each of the given CWEs using a 6-point scale (``Strongly Agree'', ``Agree'', ``Somehow Agree'', ``Somehow Disagree'', ``Disagree'', and ``Strongly Disagree''). As illustrated in Fig.~\ref{fig:surv_understand}, CWE-20 and  CWE-404 (\textit{`Improper Resource Shutdown or Release'}) were, on average, considered the weaknesses SATs help understand the most, whereas CWE-362 (\textit{`Race Condition'}) received the highest level of disagreement among participants. This same pattern was observed regarding the role of SATs in providing meaningful feedback for fixing these weaknesses (Fig.~\ref{fig:surv_fix}).

\subsubsection{SSATD-Aided Security Analysis} \label{sec:ssatd_aided}
Around 31\% of participants reported relying \textit{often} on the security pointers contained in SSATD artefacts to complement the output of SATs, regardless of the type of weakness being analysed. Another 21\% indicated doing so \textit{sometimes}, while 6\% reported doing it \textit{always}. Interestingly, although 9\% of respondents stated that they \textit{rarely} leverage SSATD instances to enhance the use of SATs, none reported \textit{never} following this practice.

\begin{figure}
    \centering
    \includegraphics[width=\linewidth]{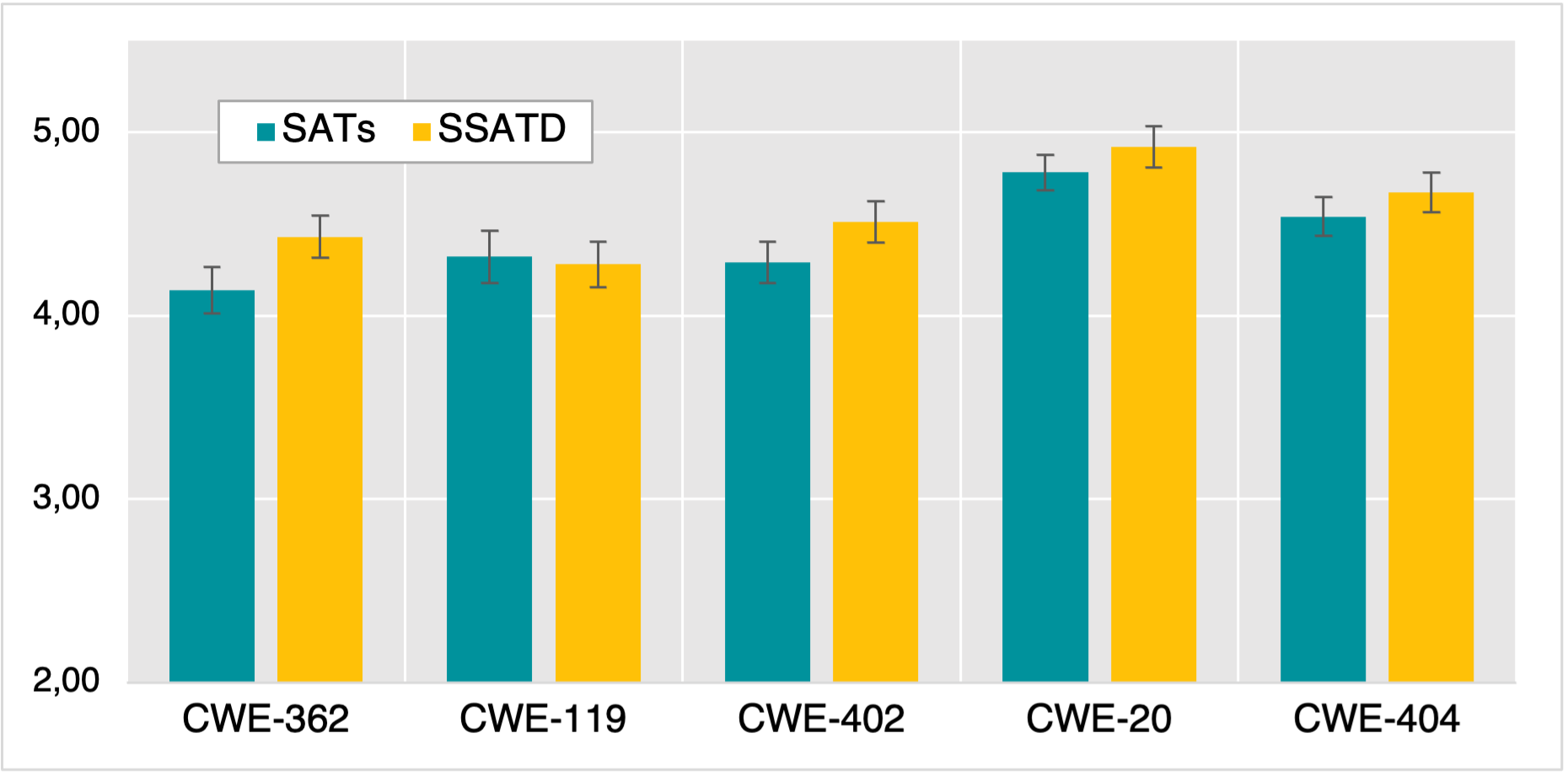}
    \caption{Perceived usefulness of SATs and SSATD comments for UNDERSTANDING different CWE types (average).}
    \label{fig:surv_understand}
\end{figure}

\begin{figure}
    \centering
    \includegraphics[width=\linewidth]{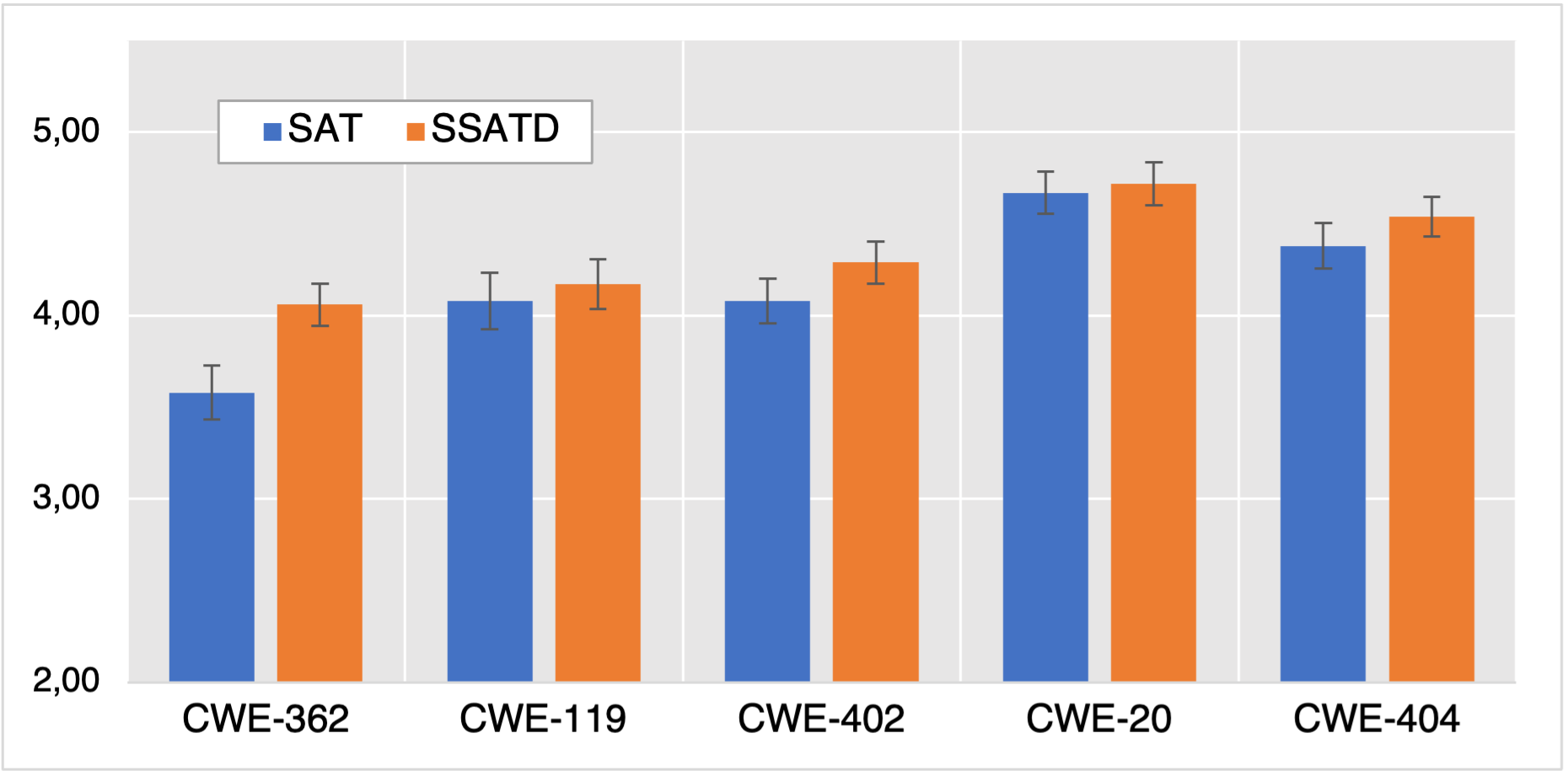}
    \caption{Perceived usefulness of SATs and SSATD comments for FIXING different CWE types (average).}
    \label{fig:surv_fix}
\end{figure}

When assessing the extent to which SSATD contributes to understanding the given security weaknesses, participants identified CWE-20 and CWE-404 as the cases where SSATD-encoded information was most helpful. Conversely, CWE-119 was deemed the weakness for which such information contributed the least to understanding (Fig.~\ref{fig:surv_understand}). Regarding the value of SSATD for fixing these weaknesses, participants once again perceived CWE-20 and CWE-404 as the most supported cases, while considering it less relevant for instances of CWE-362 (Fig.~\ref{fig:surv_fix}).

\begin{table}[t]
\centering\small
\caption{Results of paired-sample \textit{t}-tests comparing the effectiveness of SATs across different CWEs.}
\label{tab:ttests}
\begin{adjustbox}{max width=\columnwidth}
\begin{tabular}{lcccccc}
\toprule
\textbf{SATs--SSATD} & \textbf{$\Delta$} & \textbf{\textit{t}(df)} & \textbf{\textit{p}} & \textbf{\textit{d}} & \textbf{95\% CI} \\
\midrule
\textbf{Understand} & & &  &  &  \\
CWE-362 & --0.292 & --1.980(71) & 0.052 & --0.233 & [--0.585,0.002] \\
CWE-119 & \phantom{--}0.042 & \phantom{--}0.335(71) & 0.738 & \phantom{--}0.040 & [--0.206,0.289] \\
CWE-402 & --0.222 & --1.601(71) & 0.114 & --0.189 & [--0.499,0.054] \\
CWE-20 & --0.139 & --1.12(71) & 0.266 & --0.132 & [--0.386,0.108] \\
CWE-404 & --0.125 & --1.013(71) & 0.315 & --0.119 & [--0,371,0.121] \\
\midrule
\textbf{Fix} & & &  &  &  \\
CWE-362 & --0.472 & --3.730(71) & \textless.001 & --0.435 & [--0.725,--0.22] \\
CWE-119 & --0.083 & --0.736(71) & 0.464 & --0.086 & [--0.309,0.142] \\
CWE-402 & --0.208 & --1.521(71) & 0.133 & --0.177 & [--0.481,0.065] \\
CWE-20 & --0.056 & --0.406(71) & 0.686 & --0.047 & [--0.328,0.217] \\
CWE-404 & --0.167 & --1.404(71) & 0.165 & --0.164 & [--0.403,0.070] \\
\bottomrule
\end{tabular}
\end{adjustbox}
\end{table}

As shown in Fig.~\ref{fig:surv_understand}, clear differences emerge when comparing participants' assessments of SATs and SSATD regarding their usefulness for understanding each CWE type. In particular, SSATD-encoded security insights were perceived as more helpful in all cases except for CWE-119. To examine whether these differences were statistically significant, we conducted paired-sample \textit{t}-tests on the responses obtained for each individual case. The results, shown in the upper part of Table~\ref{tab:ttests}, indicate that such differences were only marginally significant for CWE-362 (\textit{p}=0.052), yielding a small effect size (\textit{d}=--0.233). We repeated this process using the values obtained for SATs and SSATD's contribution to fixing the given types of weaknesses (Fig.~\ref{fig:surv_fix}). This time, SSATD-encoded security information was perceived as more useful across all the assessed CWE variants. Nevertheless, as reported in the lower part of Table~\ref{tab:ttests}, CWE-362 was the only case showing a statistically significant difference (\textit{p}\textless 0.001), yielding a medium effect size (\textit{d}=--0.435).

\begin{tcolorbox}[colback=gray!20, boxsep=1pt]
\textbf{Findings 3}: Overall, study participants deemed the security information encoded in SSATD artefacts valuable for understanding and fixing weaknesses identified by SATs, particularly for addressing instances of CWE-362.
\end{tcolorbox}

\subsubsection{SSATD-Specific contributions}\label{sec:ssatd_contr} We observed that, in general, study participants considered SSATD-encoded information valuable for gaining insights into (i) the \textit{type}, (ii) the \textit{impact}, and (iii) the \textit{affected software components} of security weaknesses detected by SATs, as well as (iv) the \textit{attack vectors} that may exploit them. In particular, around 98.6\% of respondents acknowledged the relevance of SSATD when assessing the type of SAT-detected CWEs, followed by 88.9\% for their impact and 87.5\% for the affected components. However, a substantial 26.4\% tended to disagree on the value of SSATD for identifying potential attack vectors. Alongside, \textbf{4 additional dimensions} emerged from participants' feedback on which \textit{other CWE characteristics} SSATD helped them understand:

\vspace{1ex}
\textbf{\textit{(i) Root Causes}}: Many participants shared that SSATD often provides insight into the origins of security issues and helps surface their contributing factors. 
\begin{quote}
\textit{P43: ``SATs might flag an issue, but code comments and commit messages can provide insight into why the issue occurred in the first place.''}
\end{quote}
\begin{quote}
\textit{P72: ``[SSATD instances] can help to identify whether an issue stems from misconfigurations, insecure coding practices, or third-party dependencies.''}
\end{quote}

\vspace{1ex}
\textbf{\textit{(ii) Context and Background}}: Some participants also reported that the security insights encoded across SSATD artefacts help clarify the background and context of SAT-detected weaknesses. Such information can range from developers’ intent to the historical evolution of the security issues.
\begin{quote}
\textit{P41: ``[SSATD] artifacts help clarify when an issue was introduced or recognised, and whether it's a known risk being accepted, a regression, or a work-in-progress fix. They also provide developer intent, which helps in distinguishing between poor practice and temporary trade-offs.''}
\end{quote}
\begin{quote}
\textit{P15: ``Context around the issue: why has it not been fixed yet? Is it super rare? Is it worth it? Context is important, and SATs don't have that.''}
\end{quote}

\vspace{1ex}
\textbf{\textit{(iii) Harms}}: We also found several references to the potential harms and business impact associated with security issues. Moreover, some participants indicated that SSATD helps them gauge the overall criticality of a weakness and envision specific incidents that might result from it.
\begin{quote}
\textit{P1: ``They also give insight into whether the problem is isolated or affects multiple components. Additionally, they help assess the severity of the issue and its potential impact on the application.''}
\end{quote}
\begin{quote}
\textit{P43: ``Security pointers in artifacts help trace which parts of the system are impacted by the issue. For example: This resource leak occurs in the authentication module when handling expired sessions.''}
\end{quote}

\vspace{1ex}
\textbf{\textit{(iv) Fixing Strategies}}: Finally, many respondents emphasised that SSATD instances can provide actionable guidance for resolving security weaknesses detected through SATs, including insights into approaches that have already been attempted.
\begin{quote}
\textit{P5: ``The intended resolution for the issue, often suggested in comments or pull requests.''	}
\end{quote}
\begin{quote}
\textit{P69: ``Sometimes these security pointers can contain mitigation strategies and provide context as well.''}
\end{quote}

\begin{tcolorbox}[colback=gray!20, boxsep=1pt]
\textbf{Findings 4}: Participants deemed SSATD-encoded information particularly valuable for assessing the type, impact, and affected components of security weaknesses identified through SATs. Furthermore, they believed it also supports understanding their broader context, potential harms, and root causes, as well as devising appropriate fixing strategies.
\end{tcolorbox}

\section{DISCUSSION AND IMPLICATIONS} \label{sec:discussion}

\subsubsection*{\textbf{Contribution to SAT-driven analysis}}

The findings gathered throughout this study suggest that SSATD-encoded information can support SAT-driven security analysis by \rev{(i) broadening the security-relevant information available to practitioners and (ii) providing further insights into the origins and prospective fixes of security weaknesses.} Although prior investigations advise combining multiple SATs to capture a broader range of weaknesses (e.g., \cite{bennett2024semgrep,esposito2024extensive}), Table~\ref{tab:total_flagged} shows that this approach contributed only one additional flagged case in our particular setting (i.e., a total of 114 SAT-detected cases, 113 of which were identified by the CWE Heuristics). Nevertheless, combining the tools expanded the number of SAT-derived CWE types to 21, with 16 identified by the CWE Heuristics and 8 by Flawfinder. Conversely, 21 SSATD-annotated cases were \rev{not flagged by the selected SATs under the evaluated configurations} and were manually mapped to 29 CWE identifiers (Table~\ref{tab:method_cwes}). The assessment of CWE pairs described in Section~\ref{met:comparison} further showed that only 6.42\% of the SAT-derived and manually mapped CWE pairs were closely related. \rev{These findings provide exploratory evidence that SSATD can meaningfully complement SAT-driven security analysis by surfacing security-relevant information and weakness categories that may not be reflected in the corresponding SAT outputs. This contribution is particularly notable when contrasted with the CWE coverage of established SATs such as Flawfinder, whose rule base covers fewer than 25 CWE categories~\cite{flawfinder}.}



Although combining multiple SATs is generally regarded as good practice, empirical evidence suggests that developers rarely do so and typically rely on a single tool~\cite{bennett2024developers}. Moreover, SAT outputs can vary substantially depending on tool configuration, including the selection, modification, or deactivation of specific rules. Nevertheless, practitioners frequently use SATs out of the box, which may limit their ability to identify certain security weaknesses \cite{bennett2024developers,vassallo2020developers}. This issue was also observed in our evaluation, as Semgrep did not flag any SSATD instances \rev{under the selected default configuration}. Our survey findings are consistent with these limitations, as participants frequently reported using security information available in SSATD artefacts alongside SAT outputs (Section~\ref{sec:ssatd_aided}). More broadly, the reported practices and perceptions reflect usability limitations of SATs documented in prior work, including difficulties interpreting and acting upon tool-generated findings (e.g., \cite{tahaei2021security,smith2020can}).

\subsubsection*{\textbf{Identification of challenging CWE types}}

A closer look into the type of security weaknesses spotted through manual SSATD assessment shows that \rev{several correspond to weakness types that prior work has found challenging to detect using traditional SATs}. For instance, a large-scale study by Lipp et al.~\cite{lipp2022empirical} covering multiple free and commercial SATs revealed that issues related to (i) the \textit{improper control of resources through their lifetime} (e.g., CWE-20) and (ii) \textit{improper check or handling of exceptional conditions} (e.g., CWE-754) are often overlooked by these tools. Furthermore, the study showed that cases of (iii) \textit{incorrect calculations} (e.g., CWE-354), (iv) \textit{insufficient control flow management} (e.g., CWE-358), and (v) \textit{improper neutralization of program input/outputs} (e.g., CWE-20) often go undetected by traditional SATs, as they do not primarily involve memory-related issues \cite{bennett2024semgrep}. 

Prior work has also shown that static approaches struggle with race conditions (CWE-362) because they often cannot determine which execution paths may run concurrently \cite{dossche2024context}. In turn, some of them produce many false positive warnings or limit their scope to very specific classes of race conditions like Double-Checked Locking~(DCL) \cite{bai2019effective}. This limitation is consistent with our survey findings, which suggest that practitioners may find SSATD particularly useful for understanding and addressing such weaknesses (Section~\ref{sec:ssatd_aided}).

\added{Although security weaknesses of a dynamic nature are traditionally addressed through techniques such as dynamic analysis or runtime testing, these approaches often require substantial setup effort, execution time, and specialized expertise \cite{ lipp2022empirical}. Conversely, our findings suggest that SSATD instances can provide \rev{readily available security cues concerning issues that may not be reflected in static analysis outputs}. In that sense, SSATD should be seen not as a substitute for dynamic analysis, but as a pragmatic complement that can help prioritize, contextualize, or motivate deeper security assessments when warranted.}


\begin{figure}[t]
\begin{tcolorbox}[colback=gray!20, boxsep=1pt,title=\textbf{PRACTICAL RECOMMENDATIONS},toptitle=2pt, bottomtitle=2pt]
\begin{enumerate}[leftmargin=10pt,label=\arabic*., itemsep=1pt,topsep=3pt]
    \item \rev{\textit{\textbf{Explore SSATD as a complementary source of CWE information:}} SSATD instances can highlight security weaknesses that may not be reflected in SAT outputs, particularly dynamic or context-dependent issues such as race conditions and resource leaks. Incorporating SSATD-derived cues could therefore broaden the security-relevant information available during SAT-driven assessments.}
    \item \rev{\textit{\textbf{Use SSATD as a source of contextual metadata for SAT warnings:}} During the survey study, participants reported using SSATD to interpret SAT findings, particularly to understand their impact, root causes, and potential remediation strategies. These findings suggest opportunities to enrich SAT warnings with contextual information derived from SSATD and improve their interpretability and actionability.}
    \item \rev{\textit{\textbf{Investigate CWE-specific integration strategies:}} The perceived value of SSATD varied across the CWE types examined in the survey. This suggests that future integration efforts could prioritize security weakness categories for which SSATD appears particularly useful, rather than adopting a uniform approach across all SAT findings.}
    \item \rev{\textit{\textbf{Consider lightweight augmentation of out-of-the-box SAT usage:}} Since practitioners rarely combine multiple SATs or customize their rule sets, SSATD-based augmentation may offer a lightweight means of complementing default SAT outputs without imposing substantial additional configuration effort on developers.}
\end{enumerate}
\end{tcolorbox}
\vspace{-3ex}
\end{figure}

\subsubsection*{\textbf{Context-aware security insights}}

As several participants pointed out, contextual information is essential for understanding and fixing SAT-detected weaknesses. Yet, many SATs provide limited contextual support, an issue that prior studies have identified as a barrier to their effective adoption \cite{nachtigall2022large,muske2016survey}. Developers' knowledge and expertise could, in principle, help address this limitation, for instance by providing feedback on false positives or contextual annotations that enhance the interpretability of the detected weaknesses \cite{nachtigall2022large}. Still, limited efforts have been made to incorporate such security expertise into the design of SATs. 
\rev{Similarly, the potential role of software artefacts in SAT-aided security analysis has received comparatively little attention, with only a few recent exceptions (e.g.,~\cite{charoenwet2024toward,diaz2024satd}). Accordingly, our findings highlight opportunities for using SSATD-encoded information to enrich SAT outputs with contextual cues and motivate future research on its systematic integration into SAT-aided security workflows.}

\section{LIMITATIONS AND THREATS TO VALIDITY} \label{sec:limitations}

To a certain extent, the results of our study are subject to limitations related to its experimental design. In particular, the following construct, external, internal, and conclusion threats \cite{WohlinEtAl2012Threats} may affect the validity of our findings and practical implications:

\textbf{Construct Validity:} As mentioned across the different paper sections, SATs cannot provide a complete, accurate overview of all security weaknesses present in code. Furthermore, their effectiveness is subject to configuration parameters that can substantially affect the number and type of the identified CWEs. We have followed standard configurations for all the assessed SATs in this study for the sake of reproducibility of our findings. However, we acknowledge that this may have influenced the detection outcomes, potentially under-representing certain CWE types or over-representing others. As a result, our findings primarily reflect the insights available under typical, out-of-the-box SAT configurations rather than the full potential of these tools. Likewise, the identification of SSATD cases (and consequently their mapping to CWE-IDs) is constrained by the use of a predefined set of security keywords, which may miss implicit references to security issues or cases in which other security terminology is used. \rev{Manual validation helped remove false-positive candidates but could not account for SSATD instances omitted during keyword-based filtering. We therefore did not attempt to quantify or compare potentially missing cases across both approaches (i.e., SATs and SSATD).}


\textbf{Internal Validity:} The manual validations and mappings performed in this study are subject to the expertise and judgment of the individuals who conducted them. To mitigate potential biases in the manual assessment of SSATD candidates, each case was independently assessed by two authors and discussed jointly afterwards. Similarly, the manual mapping of SSATD instances to CWE-IDs was first performed by one author and then closely validated by another with extensive security expertise. We also sought to bring some degree of consistency to these assessments by establishing general guidelines for the authors to follow. Still, given the size and complexity of MITRE's CWE database, a completely bias-free mapping is unattainable, and some subjectivity in assigning CWE categories to SSATD instances may persist despite our validation efforts. Likewise, we acknowledge that our interpretations of some survey responses may be influenced by individual perspectives, particularly when extracting additional SSATD-specific contributions (Section~\ref{sec:ssatd_contr}) from participants' free text answers. To reduce subjectivity, \rev{these contributions were carefully reviewed and validated jointly by the authors}.

The recruitment process followed in the survey study also suffers from limitations. Particularly, it is known that prospective participants recruited via crowd-sourcing platforms may not always meet the study eligibility criteria (e.g., some of them may claim to have prior programming experience when, in fact, they do not). We have addressed this issue by (i) using Prolific's in-built qualification features and (ii) running a screening questionnaire before the main survey, as suggested when conducting studies of this nature (see \cite{tahaei2022recruiting}). Additionally, we introduced attention questions to spot random answers from dishonest participants. Overall, the detailed answers observed in the open-ended question increased our confidence that the vast majority of participants had the right level of expertise.


\textbf{External Validity:} Although the analysed reference dataset is relatively large (3,279 SATD-annotated functions), it is limited to a small number of C/C++ projects. Furthermore, our SSATD dataset comprises only 135 instances, which, together with the limited number of evaluated SATs, may constrain the generalizability of our results. \rev{We therefore interpret the findings as exploratory and contextualize them against prior literature rather than generalizing them to other project populations, programming languages, or SAT configurations}. Generalizability threats also arise from the survey study, given the size and composition of the studied sample. On the one hand, because of its size, we cannot generalise the study results to the entire community of security practitioners and SAT users. In fact, it can only account for the experiences of those who are, coincidentally, active Prolific participants. In turn, the insights we gained should be seen as preliminary and motivate further research in this area. On the other hand, because of its demographic composition, we may have overlooked the perspectives of underrepresented gender groups (e.g., women and gender-diverse people). Hence, future investigations should target larger and more gender-diverse samples for the sake of generalizability.

\textbf{Conclusion Validity:} Finally, the frequency and distribution of identified CWEs may have been influenced by the specific SAT tools used for this study. Similarly, the CWE characterization of SSATD instances depends entirely on the judgments and perspectives of the authors involved. Thus, despite our efforts to maintain objectivity in both manual and SAT-aided assessments, results could differ if alternative tools, configurations, or evaluators were employed. \rev{Moreover, several differences observed in the perceived usefulness of SATs and SSATD (Section~\ref{sec:ssatd_aided}) were not statistically significant and should therefore be interpreted as descriptive trends rather than conclusive differences.}

\section{CONCLUSION AND FUTURE WORK} \label{sec:conclusion}

The evidence gathered through our mixed-methods study \rev{suggests} that SSATD can offer complementary value to SAT-driven security analysis. On the one hand, the dataset study gives insights into the types of security weaknesses reflected in SSATD comments. As discussed in Section~\ref{sec:discussion}, these include race conditions and resource leaks, which the selected SATs did not flag under the evaluated configurations and which are often challenging to infer through static analysis due to their dynamic or context-dependent nature. The survey findings further suggest that practitioners use SSATD to understand relevant aspects of security weaknesses, including their impact, root causes, affected components, and potential fixes. \rev{Nevertheless, SSATD and SAT outputs remain loosely coupled in practice, with their combined use depending primarily on developers' individual efforts and interpretations.}


Future research efforts should focus on the automatic identification of SSATD instances and their systematic integration into SAT-driven security assessments. In this vein, Charoenwet et al.~\cite{charoenwet2024toward} proposed a deep learning model capable of semi-automatically detecting references to security issues in code review comments. Such an approach could, in principle, be adapted to SSATD identification through transfer learning, but it would require a larger and more diverse dataset. \rev{Large Language Models (LLMs) such as GPT-5 could also support this process by assisting in the contextual interpretation of developer comments and improving the precision of SSATD classification \cite{diaz2024satd}. Nevertheless, emerging evidence suggests that generative AI tools may change how developers comment and document code~\cite{ebiwonjumi2026generative,sergeyuk2025using}. Future investigations should therefore examine how these changes affect the expression and frequency of SATD. They should also explore how LLMs can document security shortcomings through SATD patterns while surfacing contextual information relevant to SAT-driven security assessments.}


\section*{Replication Package}\label{sec:replication}

Dataset, survey instruments, consent forms, and study results are available at \url{https://doi.org/10.5281/zenodo.18460401}

\section*{Acknowledgments}
Moritz Mock is funded by the European Union- Next Generation EU, Mission 4 Component 1 CUP I52B23000570003. Moritz Mock and Barbara Russo thank the project CyberSecurity Laboratory no. EFRE1039 under the 2023 EFRE/FESR program. This work was supported by the Deutsche Forschungsgemeinschaft (DFG, German Research Foundation) under project number 575580343, ``Developer-Centred Management of Self-Admitted Technical Debt for Security'' (DevSSATD).

\begin{appendices}

\section{Survey Questions} \label{appendix}

The survey questions described in Section~\ref{sec:survey_study} are listed below. Note that \textbf{Q1}, \textbf{Q2}, \textbf{Q3}, \textbf{Q5}, and \textbf{Q6} iterate on each of the five given CWE types (\texttt{CWE-XYZ}), namely \texttt{CWE-362}, \texttt{CWE-119}, \texttt{CWE-402}, \texttt{CWE-20}, and \texttt{CWE-404}. Hence, they are asked five times. As illustrated in Fig~\ref{fig:vignete}, each of these questions is complemented with a short description of the corresponding CWE type. The complete survey instrument, including demographic questions and the short technical questionnaire used to pre-screen the participants (Section~\ref{sec:population}), is available in the paper's \hyperref[sec:replication]{Replication Rackage}.

\begin{itemize}[leftmargin=*]
 \item \textbf{Q1}$^*$: Please indicate how often you \texttt{IDENTIFY} [\texttt{CWE-XYZ}] with the help of SATs.\\[0.5ex]\underline{Answer options}: \textit{Never, Rarely, Sometimes, Often, Always}.\vspace{1ex}
 \item \textbf{Q2}$^*$: ``SATs provide enough information to \texttt{UNDERSTAND} [\texttt{CWE-XYZ}] in source code.''
  \\[0.5ex]\underline{Answer options}: \textit{Completely Disagree, Disagree, Somehow Disagree, Somehow Agree, Agree, Completely Agree}.\vspace{1ex}
  \item \textbf{Q3}$^*$: ``SATs provide enough information to \texttt{FIX} [\texttt{CWE-XYZ}] in source code.''
 \\[0.5ex]\underline{Answer options}: \textit{Completely Disagree, Disagree, Somehow Disagree, Somehow Agree, Agree, Completely Agree}.\vspace{1ex}
 \item \textbf{Q4}: How often do you use security pointers contained in software artefacts to complement the output of SATs?
 \\[0.5ex]\underline{Answer options}: \textit{Never, Rarely, Sometimes, Often, Always}.\vspace{1ex}
 \item \textbf{Q5}$^*$: ``The security pointers contained inside software artefacts help me \texttt{UNDERSTAND} [\texttt{CWE-XYZ}]''
  \\[0.5ex]\underline{Answer options}: \textit{Completely Disagree, Disagree, Somehow Disagree, Somehow Agree, Agree, Completely Agree}.\vspace{1ex}
  \item \textbf{Q6}$^*$: ``The security pointers contained inside software artefacts help me \texttt{FIX} [\texttt{CWE-XYZ}]''
 \\[0.5ex]\underline{Answer options}: \textit{Completely Disagree, Disagree, Somehow Disagree, Somehow Agree, Agree, Completely Agree}.\vspace{1ex}
 \item \textbf{Q7}: ``The security pointers contained inside software artefacts help me to understand the \texttt{TYPE} of issues spotted by SATs.''
  \\[0.5ex]\underline{Answer options}: \textit{Completely Disagree, Disagree, Somehow Disagree, Somehow Agree, Agree, Completely Agree}.\vspace{1ex}
 \item \textbf{Q8}: ``The security pointers contained inside software artefacts help me to understand the \texttt{IMPACT} of issues spotted by SATs.''
  \\[0.5ex]\underline{Answer options}: \textit{Completely Disagree, Disagree, Somehow Disagree, Somehow Agree, Agree, Completely Agree}.\vspace{1ex}
 \item \textbf{Q9}: ``The security pointers contained inside software artefacts help me to understand the \texttt{ATTACK VECTORS} of issues spotted by SATs.''
  \\[0.5ex]\underline{Answer options}: \textit{Completely Disagree, Disagree, Somehow Disagree, Somehow Agree, Agree, Completely Agree}.\vspace{1ex}
 \item \textbf{Q10}: ``The security pointers contained inside software artefacts help me to understand the \texttt{AFFECTED COMPONENTS} of issues spotted by SATs.''
 \\[0.5ex]\underline{Answer options}: \textit{Completely Disagree, Disagree, Somehow Disagree, Somehow Agree, Agree, Completely Agree}.\vspace{1ex}
 \item \textbf{Q11}: What other characteristics can the security pointers contained in software artifacts help you understand?
 \\[0.5ex]\underline{Answer options}: FREE TEXT.
\end{itemize}
\footnotesize (*) This question is asked for each CWE type.
\end{appendices}

\bibliographystyle{elsarticle-num} 
\bibliography{references}

@inproceedings{MockEtAl2024Dataset,
    author = {Mock, Moritz and Melegati, Jorge and Kretschmann, Max and Diaz Ferreyra, Nicolas E. and Russo, Barbara},
    title = {MADE-WIC: Multiple Annotated Datasets for Exploring Weaknesses In Code},
    year = {2024},
    isbn = {9798400712487},
    publisher = {Association for Computing Machinery},
    address = {New York, NY, USA},
    url = {https://doi.org/10.1145/3691620.3695348},
    doi = {10.1145/3691620.3695348},
    booktitle = {Proceedings of the 39th IEEE/ACM International Conference on Automated Software Engineering},
    pages = {2346–2349},
    numpages = {4},
    location = {Sacramento, CA, USA},
    series = {ASE '24}
}

@inproceedings{RussoEtAl2022Weaksatd,
    author = {Russo, Barbara and Camilli, Matteo and Mock, Moritz},
    title = {WeakSATD: detecting weak self-admitted technical debt},
    year = {2022},
    isbn = {9781450393034},
    publisher = {Association for Computing Machinery},
    address = {New York, NY, USA},
    url = {https://doi.org/10.1145/3524842.3528469},
    doi = {10.1145/3524842.3528469},
    booktitle = {Proceedings of the 19th International Conference on Mining Software Repositories},
    pages = {448–453},
    numpages = {6},
    location = {Pittsburgh, Pennsylvania},
    series = {MSR '22}
}

@article{ZhouEtAl2019Devign,
    author = {Zhou, Yaqin and Liu, Shangqing and Siow, Jingkai and Du, Xiaoning and Liu, Yang},
    booktitle = {Advances in Neural Information Processing Systems},
    editor = {H. Wallach and H. Larochelle and A. Beygelzimer and F. d\textquotesingle Alch\'{e}-Buc and E. Fox and R. Garnett},
    pages = {},
    publisher = {Curran Associates, Inc.},
    title = {Devign: Effective Vulnerability Identification by Learning Comprehensive Program Semantics via Graph Neural Networks},
    url = {https://proceedings.neurips.cc/paper_files/paper/2019/file/49265d2447bc3bbfe9e76306ce40a31f-Paper.pdf},
    volume = {32},
    year = {2019}
}

@inproceedings{FanEtAl2020Big-Vul,
    author = {Fan, Jiahao and Li, Yi and Wang, Shaohua and Nguyen, Tien N.},
    title = {A C/C++ Code Vulnerability Dataset with Code Changes and CVE Summaries},
    year = {2020},
    isbn = {9781450375177},
    publisher = {Association for Computing Machinery},
    address = {New York, NY, USA},
    url = {https://doi.org/10.1145/3379597.3387501},
    doi = {10.1145/3379597.3387501},
    booktitle = {Proceedings of the 17th International Conference on Mining Software Repositories},
    pages = {508–512},
    numpages = {5},
    location = {Seoul, Republic of Korea},
    series = {MSR '20}
}

@article{CroftEtAl2022SecI,
  title={An empirical study of developers’ discussions about security challenges of different programming languages},
  author={Croft, Roland and Xie, Yongzheng and Zahedi, Mansooreh and Babar, M Ali and Treude, Christoph},
  journal={Empirical Software Engineering},
  volume={27},
  pages={1--52},
  year={2022},
  publisher={Springer},
  url={https://doi.org/10.1007/s10664-021-10054-w}
}

@inproceedings{PotdarShihab2014SATD,
  author={Potdar, Aniket and Shihab, Emad},
  booktitle={2014 IEEE International Conference on Software Maintenance and Evolution}, 
  title={An Exploratory Study on Self-Admitted Technical Debt}, 
  year={2014},
  volume={},
  number={},
  pages={91-100},
  url={https://doi.org/10.1109/ICSME.2014.31}
}

@article{GuoEtAl2021MAT,
    author = {Guo, Zhaoqiang and Liu, Shiran and Liu, Jinping and Li, Yanhui and Chen, Lin and Lu, Hongmin and Zhou, Yuming},
    title = {How Far Have We Progressed in Identifying Self-admitted Technical Debts? A Comprehensive Empirical Study},
    year = {2021},
    issue_date = {October 2021},
    publisher = {Association for Computing Machinery},
    address = {New York, NY, USA},
    volume = {30},
    number = {4},
    issn = {1049-331X},
    url = {https://doi.org/10.1145/3447247},
    doi = {10.1145/3447247},
    journal = {ACM Trans. Softw. Eng. Methodol.},
    month = jul,
    articleno = {45},
    numpages = {56}
}

@misc{bandit,
  author      = {PyCQA and its OS community},
  title       = {Bandit},
  url         = {https://github.com/PyCQA/bandit},
  note        = {Accessed on {{2025-03-06}}}
}

@misc{codechecker,
  author      = {Ericsson and its OS community},
  title       = {CodeChecker},
  url         = {https://github.com/Ericsson/codechecker},
  note        = {Accessed on {{2025-03-06}}}
}

@misc{codeql,
  author      = {GitHub},
  title       = {CodeQL},
  url         = {https://codeql.github.com},
  note        = {Accessed on {{2025-03-06}}}
}

@misc{cppcheck,
  author      = {Marjamäki, Daniel and others},
  title       = {Cppcheck: A tool for static C/C++ code analysis},
  url         = {http://cppcheck.sourceforge.net},
  note        = {Accessed on {{2025-03-06}}}
}

@misc{flawfinder,
  author      = {Wheeler, David A.},
  title       = {Flawfinder},
  url         = {https://dwheeler.com/flawfinder/flawfinder.pdf},
  note        = {Accessed on {{2025-03-06}}}
}

@misc{semgrep,
  author      = {Semgrep},
  title       = {Semgrep},
  url         = {https://semgrep.dev},
  note        = {Accessed on {{2025-03-06}}}
}

@inproceedings{ImprotaEtAl2025semgrep,
  author={Improta, Cristina and Tufano, Rosalia and Liguori, Pietro and Cotroneo, Domenico and Bavota, Gabriele},
  booktitle={2025 IEEE/ACM 33rd International Conference on Program Comprehension (ICPC)}, 
  title={Quality In, Quality Out: Investigating Training Data's Role in AI Code Generation}, 
  year={2025},
  volume={},
  number={},
  pages={454-465},
  url={https://doi.org/10.1109/ICPC66645.2025.00056}
}

@inproceedings{esposito2024extensive,
    author = {Esposito, Matteo and Falaschi, Valentina and Falessi, Davide},
    title = {An Extensive Comparison of Static Application Security Testing Tools},
    year = {2024},
    isbn = {9798400717017},
    publisher = {Association for Computing Machinery},
    address = {New York, NY, USA},
    url = {https://doi.org/10.1145/3661167.3661199},
    doi = {10.1145/3661167.3661199},
    booktitle = {Proceedings of the 28th International Conference on Evaluation and Assessment in Software Engineering},
    pages = {69–78},
    numpages = {10},
    location = {Salerno, Italy},
    series = {EASE '24}
}

@inproceedings{croft2021empirical,
author = {Croft, Roland and Newlands, Dominic and Chen, Ziyu and Babar, M. Ali},
title = {An Empirical Study of Rule-Based and Learning-Based Approaches for Static Application Security Testing},
year = {2021},
isbn = {9781450386654},
publisher = {Association for Computing Machinery},
address = {New York, NY, USA},
url = {https://doi.org/10.1145/3475716.3475781},
doi = {10.1145/3475716.3475781},
booktitle = {Proceedings of the 15th ACM / IEEE International Symposium on Empirical Software Engineering and Measurement (ESEM)},
articleno = {8},
numpages = {12},
location = {Bari, Italy},
series = {ESEM '21}
}

@inproceedings{tahaei2021security,
    author = {Tahaei, Mohammad and Vaniea, Kami and Beznosov, Konstantin (Kosta) and Wolters, Maria K},
    title = {Security Notifications in Static Analysis Tools: Developers’ Attitudes, Comprehension, and Ability to Act on Them},
    year = {2021},
    isbn = {9781450380966},
    publisher = {Association for Computing Machinery},
    address = {New York, NY, USA},
    url = {https://doi.org/10.1145/3411764.3445616},
    doi = {10.1145/3411764.3445616},
    booktitle = {Proceedings of the 2021 CHI Conference on Human Factors in Computing Systems},
    articleno = {691},
    numpages = {17},
    location = {Yokohama, Japan},
    series = {CHI '21}
}

@inproceedings{bennett2024developers,
author = {Bennett, Gareth and Hall, Tracy and Counsell, Steve and Winter, Emily and Shippey, Thomas},
title = {Do Developers Use Static Application Security Testing (SAST) Tools Straight Out of the Box? A large-scale Empirical Study},
year = {2024},
isbn = {9798400710476},
publisher = {Association for Computing Machinery},
address = {New York, NY, USA},
url = {https://doi.org/10.1145/3674805.3690750},
doi = {10.1145/3674805.3690750},
booktitle = {Proceedings of the 18th ACM/IEEE International Symposium on Empirical Software Engineering and Measurement},
pages = {454–460},
numpages = {7},
location = {Barcelona, Spain},
series = {ESEM '24}
}

@article{vassallo2020developers,
  title={How developers engage with static analysis tools in different contexts},
  author={Vassallo, Carmine and Panichella, Sebastiano and Palomba, Fabio and Proksch, Sebastian and Gall, Harald C and Zaidman, Andy},
  journal={Empirical Software Engineering},
  volume={25},
  number={2},
  pages={1419--1457},
  year={2020},
  publisher={Springer},
  url={https://doi.org/10.1007/s10664-019-09750-5}
}

@inproceedings{bennett2024semgrep,
    author = {Bennett, Gareth and Hall, Tracy and Winter, Emily and Counsell, Steve},
    title = {Semgrep*: Improving the Limited Performance of Static Application Security Testing (SAST) Tools},
    year = {2024},
    isbn = {9798400717017},
    publisher = {Association for Computing Machinery},
    address = {New York, NY, USA},
    url = {https://doi.org/10.1145/3661167.3661262},
    doi = {10.1145/3661167.3661262},
    booktitle = {Proceedings of the 28th International Conference on Evaluation and Assessment in Software Engineering},
    pages = {614–623},
    numpages = {10},
    location = {Salerno, Italy},
    series = {EASE '24}
}

@article{yedida2023find,
  author={Yedida, Rahul and Kang, Hong Jin and Tu, Huy and Yang, Xueqi and Lo, David and Menzies, Tim},
  journal={IEEE Transactions on Software Engineering}, 
  title={How to Find Actionable Static Analysis Warnings: A Case Study With FindBugs}, 
  year={2023},
  volume={49},
  number={4},
  pages={2856-2872},
  url={https://doir.org/10.1109/TSE.2023.3234206}
}

@article{nam2019bug,
    title = {A bug finder refined by a large set of open-source projects},
    journal = {Information and Software Technology},
    volume = {112},
    pages = {164-175},
    year = {2019},
    issn = {0950-5849},
    url = {https://doi.org/10.1016/j.infsof.2019.04.014},
    urlIGNORE = {https://www.sciencedirect.com/science/article/pii/S0950584919300977},
    author = {Jaechang Nam and Song Wang and Yuan Xi and Lin Tan}
}

@inproceedings{nachtigall2022large,
    author = {Nachtigall, Marcus and Schlichtig, Michael and Bodden, Eric},
    title = {A large-scale study of usability criteria addressed by static analysis tools},
    year = {2022},
    isbn = {9781450393799},
    publisher = {Association for Computing Machinery},
    address = {New York, NY, USA},
    url = {https://doi.org/10.1145/3533767.3534374},
    doi = {10.1145/3533767.3534374},
    booktitle = {Proceedings of the 31st ACM SIGSOFT International Symposium on Software Testing and Analysis},
    pages = {532–543},
    numpages = {12},
    location = {Virtual, South Korea},
    series = {ISSTA 2022}
}

@inproceedings{muske2016survey,
  author={Muske, Tukaram and Serebrenik, Alexander},
  booktitle={2016 IEEE 16th International Working Conference on Source Code Analysis and Manipulation (SCAM)}, 
  title={Survey of Approaches for Handling Static Analysis Alarms}, 
  year={2016},
  volume={},
  number={},
  pages={157-166},
  url={https://doi.org/10.1109/SCAM.2016.25}
}

@inproceedings{braz2022less,
    author = {Braz, Larissa and Aeberhard, Christian and \c{C}alikli, G\"{u}l and Bacchelli, Alberto},
    title = {Less is more: supporting developers in vulnerability detection during code review},
    year = {2022},
    isbn = {9781450392211},
    publisher = {Association for Computing Machinery},
    address = {New York, NY, USA},
    url = {https://doi.org/10.1145/3510003.3511560},
    doi = {10.1145/3510003.3511560},
    booktitle = {Proceedings of the 44th International Conference on Software Engineering},
    pages = {1317–1329},
    numpages = {13},
    location = {Pittsburgh, Pennsylvania},
    series = {ICSE '22}
}

@article{elder2022really,
  title={Do i really need all this work to find vulnerabilities? an empirical case study comparing vulnerability detection techniques on a java application},
  author={Elder, Sarah and Zahan, Nusrat and Shu, Rui and Metro, Monica and Kozarev, Valeri and Menzies, Tim and Williams, Laurie},
  journal={Empirical Software Engineering},
  volume={27},
  number={6},
  pages={154},
  year={2022},
  publisher={Springer},
  url={https://doi.org/10.1007/s10664-022-10179-6}
}

@inproceedings{smith2020can,
author = {Justin Smith and Lisa Nguyen Quang Do and Emerson Murphy-Hill},
title = {Why Can{\textquoteright}t Johnny Fix Vulnerabilities: A Usability Evaluation of Static Analysis Tools for Security},
booktitle = {Sixteenth Symposium on Usable Privacy and Security (SOUPS 2020)},
year = {2020},
isbn = {978-1-939133-16-8},
pages = {221--238},
url = {https://www.usenix.org/conference/soups2020/presentation/smith},
publisher = {USENIX Association},
month = aug
}

@inproceedings{rindell2019sec,
  author={Rindell, Kalle and Holvitie, Johannes},
  booktitle={2019 International Conference on Cyber Security and Protection of Digital Services (Cyber Security)}, 
  title={Security Risk Assessment and Management as Technical Debt}, 
  year={2019},
  volume={},
  number={},
  pages={1-8},
  url={https://doi.org/10.1109/CyberSecPODS.2019.8885100}
}

@inproceedings{rindell2019managing,
author = {Rindell, Kalle and Bernsmed, Karin and Jaatun, Martin Gilje},
title = {Managing Security in Software: Or: How I Learned to Stop Worrying and Manage the Security Technical Debt},
year = {2019},
isbn = {9781450371643},
publisher = {Association for Computing Machinery},
address = {New York, NY, USA},
url = {https://doi.org/10.1145/3339252.3340338},
doi = {10.1145/3339252.3340338},
booktitle = {Proceedings of the 14th International Conference on Availability, Reliability and Security},
articleno = {60},
numpages = {8},
location = {Canterbury, CA, United Kingdom},
series = {ARES '19}
}

@inproceedings{coetzer2024managing,
  title={Managing cyber security debt: strategies for identification, prioritisation, and mitigation},
  author={Coetzer, Christo and Leenen, Louise},
  booktitle={International Conference on Cyber Warfare and Security},
  pages={439--446},
  year={2024},
  organization={Academic Conferences International Limited}
}

@inproceedings{diaz2024satd,
    author = {D\'{\i}az Ferreyra, Nicol\'{a}s E. and Shahin, Mojtaba and Zahedi, Mansooreh and Quadri, Sodiq and Scandariato, Riccardo},
    title = {What Can Self-Admitted Technical Debt Tell Us About Security? A Mixed-Methods Study},
    year = {2024},
    isbn = {9798400705878},
    publisher = {Association for Computing Machinery},
    address = {New York, NY, USA},
    url = {https://doi.org/10.1145/3643991.3644909},
    doi = {10.1145/3643991.3644909},
    booktitle = {Proceedings of the 21st International Conference on Mining Software Repositories},
    pages = {704–715},
    numpages = {12},
    location = {Lisbon, Portugal},
    series = {MSR '24}
}

@article{freire2025comprehensive,
    author = {Freire, S\'{a}vio and Pacheco, Alexia and Rios, Nicolli and P\'{e}rez, Boris and Castellanos, Camilo and Correal, Dar\'{\i}o and Rama\v{c}, Robert and Mandi\'{c}, Vladimir and Tau\v{s}an, Neboj\v{s}a and L\'{o}pez, Gustavo and Mendon\c{c}a, Manoel and Falessi, Davide and Izurieta, Clemente and Seaman, Carolyn and Sp\'{\i}nola, Rodrigo},
    title = {A Comprehensive View on TD Prevention Practices and Reasons for Not Preventing It},
    year = {2024},
    issue_date = {September 2024},
    publisher = {Association for Computing Machinery},
    address = {New York, NY, USA},
    volume = {33},
    number = {7},
    issn = {1049-331X},
    url = {https://doi.org/10.1145/3674727},
    doi = {10.1145/3674727},
    month = sep,
    articleno = {178},
    numpages = {44},
    journal = {Transactions on Software Engineering and Methodology (TOSEM)}
}

@inproceedings{oyetoyan2018myths,
    author="Oyetoyan, Tosin Daniel
    and Milosheska, Bisera
    and Grini, Mari
    and Soares Cruzes, Daniela",
    editor="Garbajosa, Juan
    and Wang, Xiaofeng
    and Aguiar, Ademar",
    title="Myths and Facts About Static Application Security Testing Tools: An Action Research at Telenor Digital",
    booktitle="Agile Processes in Software Engineering and Extreme Programming",
    year="2018",
    publisher="Springer International Publishing",
    address="Cham",
    pages="86--103",
    url={https://doi.org/10.1007/978-3-319-91602-6_6}
}

@INPROCEEDINGS{yang2019towards,
  author={Yang, Jinqiu and Tan, Lin and Peyton, John and A Duer, Kristofer},
  booktitle={2019 IEEE/ACM 41st International Conference on Software Engineering: Software Engineering in Practice (ICSE-SEIP)}, 
  title={Towards Better Utilizing Static Application Security Testing}, 
  year={2019},
  volume={},
  number={},
  pages={51-60},
  url={https://doi.org/10.1109/ICSE-SEIP.2019.00014}
}

@INPROCEEDINGS{chembakottu2025usab,
  author={Chembakottu, Bhagya and Robillard, Martin P.},
  booktitle={2025 IEEE International Conference on Software Analysis, Evolution and Reengineering - Companion (SANER-C)}, 
  title={Usability of Static Application Security Testing Workflows}, 
  year={2025},
  url={https://www.cs.mcgill.ca/~martin/papers/2025-msr4ps.pdf}
}

@inproceedings{romano2022static,
    author = {Romano, Simone and Zampetti, Fiorella and Baldassarre, Maria Teresa and Di Penta, Massimiliano and Scanniello, Giuseppe},
    title = {Do Static Analysis Tools Affect Software Quality when Using Test-driven Development?},
    year = {2022},
    isbn = {9781450394277},
    publisher = {Association for Computing Machinery},
    address = {New York, NY, USA},
    url = {https://doi.org/10.1145/3544902.3546233},
    doi = {10.1145/3544902.3546233},
    booktitle = {Proceedings of the 16th ACM / IEEE International Symposium on Empirical Software Engineering and Measurement},
    pages = {80–91},
    numpages = {12},
    location = {Helsinki, Finland},
    series = {ESEM '22}
}

@article{charoenwet2024toward,
  title={Toward effective secure code reviews: an empirical study of security-related coding weaknesses},
  author={Charoenwet, Wachiraphan and Thongtanunam, Patanamon and Pham, Van-Thuan and Treude, Christoph},
  journal={Empirical Software Engineering},
  volume={29},
  number={4},
  pages={88},
  year={2024},
  publisher={Springer},
  url={https://doi.org/10.1007/s10664-024-10496-y}
}

@article{zampetti2022using,
  title={Using code reviews to automatically configure static analysis tools},
  author={Zampetti, Fiorella and Mudbhari, Saghan and Arnaoudova, Venera and Di Penta, Massimiliano and Panichella, Sebastiano and Antoniol, Giuliano},
  journal={Empirical Software Engineering},
  volume={27},
  number={1},
  pages={28},
  year={2022},
  publisher={Springer},
  url={https://doi.org/10.1007/s10664-021-10076-4}
}

@inproceedings{obrien202223,
    author = {OBrien, David and Biswas, Sumon and Imtiaz, Sayem and Abdalkareem, Rabe and Shihab, Emad and Rajan, Hridesh},
    title = {23 shades of self-admitted technical debt: an empirical study on machine learning software},
    year = {2022},
    isbn = {9781450394130},
    publisher = {Association for Computing Machinery},
    address = {New York, NY, USA},
    url = {https://doi.org/10.1145/3540250.3549088},
    doi = {10.1145/3540250.3549088},
    booktitle = {Proceedings of the 30th ACM Joint European Software Engineering Conference and Symposium on the Foundations of Software Engineering},
    pages = {734–746},
    numpages = {13},
    location = {Singapore, Singapore},
    series = {ESEC/FSE 2022}
}

@inproceedings{ojameruaye2016sustainability,
    author = {Ojameruaye, Bendra and Bahsoon, Rami and Duboc, Leticia},
    title = {Sustainability debt: a portfolio-based approach for evaluating sustainability requirements in architectures},
    year = {2016},
    isbn = {9781450342056},
    publisher = {Association for Computing Machinery},
    address = {New York, NY, USA},
    url = {https://doi.org/10.1145/2889160.2889218},
    doi = {10.1145/2889160.2889218},
    booktitle = {Proceedings of the 38th International Conference on Software Engineering Companion},
    pages = {543–552},
    numpages = {10},
    location = {Austin, Texas},
    series = {ICSE '16}
}

@inproceedings{sutoyo2025tracing,
  author={Sutoyo, Edi and Avgeriou, Paris and Capiluppi, Andrea},
  booktitle={2025 IEEE 22nd International Conference on Software Architecture (ICSA)}, 
  title={Tracing the Lifecycle of Architecture Technical Debt in Software Systems: A Dependency Approach}, 
  year={2025},
  volume={},
  number={},
  pages={199-209},
  url={https://doi.org/10.1109/ICSA65012.2025.00028}
}

@INPROCEEDINGS{gu2024satd,
  author={Gu, Hao and Zhang, Shichao and Huang, Qiao and Liao, Zhifang and Liu, Jiakun and Lo, David},
  booktitle={2024 IEEE International Conference on Software Analysis, Evolution and Reengineering (SANER)}, 
  title={Self-Admitted Technical Debts Identification: How Far Are We?}, 
  year={2024},
  volume={},
  number={},
  pages={804-815},
  url={https://doi.org/10.1109/SANER60148.2024.00087}
}

@article{li2025impact,
    author = {Li, Qingyuan and Yin, Zhixin and Yang, Yaopeng and Li, Chuanyi and Shen, Zongwen and Ge, Jidong and Zhong, Wenkang and Luo, Bin and Ng, Vincent},
    title = {IMPACT: Identifying and Classifying Multiple Sourced and Categorized Self-Admitted Technical Debts},
    year = {2025},
    publisher = {Association for Computing Machinery},
    address = {New York, NY, USA},
    issn = {1049-331X},
    url = {https://doi.org/10.1145/3747180},
    doi = {10.1145/3747180},
    note = {Just Accepted},
    journal = {ACM Trans. Softw. Eng. Methodol.},
    month = jul
}

@article{li2023automatic,
  title={Automatic identification of self-admitted technical debt from four different sources},
  author={Li, Yikun and Soliman, Mohamed and Avgeriou, Paris},
  journal={Empirical Software Engineering},
  volume={28},
  number={3},
  pages={65},
  year={2023},
  publisher={Springer},
  url={https://doi.org/10.1007/s10664-023-10297-9}
}

@article{hassan2025characterising,
    title = {Characterising reproducibility debt in scientific software: A systematic literature review},
    journal = {Journal of Systems and Software},
    volume = {222},
    pages = {112327},
    year = {2025},
    issn = {0164-1212},
    url = {https://doi.org/10.1016/j.jss.2024.112327},
    urlIGNORE = {https://www.sciencedirect.com/science/article/pii/S0164121224003716},
    author = {Zara Hassan and Christoph Treude and Michael Norrish and Graham Williams and Alex Potanin}
}

@article{sheikhaei2024empirical,
  title={An empirical study on the effectiveness of large language models for satd identification and classification},
  author={Sheikhaei, Mohammad Sadegh and Tian, Yuan and Wang, Shaowei and Xu, Bowen},
  journal={Empirical Software Engineering},
  volume={29},
  number={6},
  pages={159},
  year={2024},
  publisher={Springer},
  url={https://doi.org/10.1007/s10664-024-10548-3}
}

@inproceedings{kruke2024defining,
    author="Kruke, Maren Maritsdatter
    and Martini, Antonio
    and Cruzes, Daniela S.
    and Iovan, Monica",
    editor="Pfahl, Dietmar
    and Gonzalez Huerta, Javier
    and Kl{\"u}nder, Jil
    and Anwar, Hina",
    title="Defining Security Debt: A Case Study Based on Practice",
    booktitle="Product-Focused Software Process Improvement",
    year="2025",
    publisher="Springer Nature Switzerland",
    address="Cham",
    pages="43--59",
    url={https://doi.org/10.1007/978-3-031-78386-9_4}
}

@article{siavvas2022technical,
  title={Technical debt as an indicator of software security risk: a machine learning approach for software development enterprises},
  author={Siavvas, Miltiadis and Tsoukalas, Dimitrios and Jankovic, Marija and Kehagias, Dionysios and Tzovaras, Dimitrios},
  journal={Enterprise Information Systems},
  volume={16},
  number={5},
  pages={1824017},
  year={2022},
  publisher={Taylor \& Francis},
  url={https://doi.org/10.1080/17517575.2020.1824017}
}

@inproceedings{izurieta2018td,
    author = {Izurieta, Clemente and Rice, David and Kimball, Kali and Valentien, Tessa},
    title = {A Position Study to Investigate Technical Debt Associated with Security Weaknesses}, 
    year = {2018}, 
    isbn = {9781450357135}, 
    publisher = {Association for Computing Machinery}, 
    address = {New York, NY, USA}, 
    url = {https://doi.org/10.1145/3194164.3194167}, 
    doi = {10.1145/3194164.3194167},
    booktitle = {Proceedings of the 2018 International Conference on Technical Debt},
    pages = {138–142},
    numpages = {5},
    location = {Gothenburg, Sweden},
    series = {TechDebt '18} 
}

@inproceedings{pepe2024taxonomy,
  author={Pepe, Federica and Zampetti, Fiorella and Mastropaolo, Antonio and Bavota, Gabriele and Di Penta, Massimiliano},
  booktitle={2024 IEEE International Conference on Software Maintenance and Evolution (ICSME)}, 
  title={A Taxonomy of Self-Admitted Technical Debt in Deep Learning Systems}, 
  year={2024},
  volume={},
  number={},
  pages={388-399},
  doi={https://doi.org/10.1109/ICSME58944.2024.00043}
}

@inproceedings{yonekura2025context,
  author={Yonekura, Miki and Kashiwa, Yutaro and Lin, Bin and Fujiwara, Kenji and Iida, Hajimu},
  booktitle={2025 IEEE/ACM 33rd International Conference on Program Comprehension (ICPC)}, 
  title={Leveraging Context Information for Self-Admitted Technical Debt Detection}, 
  year={2025},
  volume={},
  number={},
  pages={01-12},
  url={https://doi.org/10.1109/ICPC66645.2025.00018}
}

@article{landis1977measurement,
  title={The measurement of observer agreement for categorical data},
  author={Landis, J Richard and Koch, Gary G},
  journal={biometrics},
  pages={159--174},
  year={1977},
  publisher={JSTOR}
}

@incollection{WohlinEtAl2012Threats,
    address = {Berlin, Heidelberg},
    author = {Wohlin, Claes and Runeson, Per and H{\"{o}}st, Martin and Ohlsson, Magnus C. and Regnell, Bj{\"{o}}rn and Wessl{\'{e}}n, Anders},
    booktitle = {Experimentation in Software Engineering},
    doi = {10.1007/978-3-642-29044-2_8},
    isbn = {9783642290442},
    pages = {89--116},
    publisher = {Springer Berlin Heidelberg},
    title = {{Planning}},
    volume = {9783642290},
    year = {2012}
}

@inproceedings{le2020puminer,
author = {Le, Triet Huynh Minh and Hin, David and Croft, Roland and Babar, M. Ali},
title = {PUMiner: Mining Security Posts from Developer Question and Answer Websites with PU Learning},
year = {2020},
isbn = {9781450375177},
publisher = {Association for Computing Machinery},
address = {New York, NY, USA},
url = {https://doi.org/10.1145/3379597.3387443},
doi = {10.1145/3379597.3387443},
booktitle = {Proceedings of the 17th International Conference on Mining Software Repositories},
pages = {350–361},
numpages = {12},
location = {Seoul, Republic of Korea},
series = {MSR '20}
}

@inproceedings{tahaei2022recruiting,
    author = {Tahaei, Mohammad and Vaniea, Kami},
    title = {Recruiting Participants With Programming Skills: A Comparison of Four Crowdsourcing Platforms and a CS Student Mailing List},
    year = {2022},
    isbn = {9781450391573},
    publisher = {Association for Computing Machinery},
    address = {New York, NY, USA},
    url = {https://doi.org/10.1145/3491102.3501957},
    doi = {10.1145/3491102.3501957},
    booktitle = {Proceedings of the 2022 CHI Conference on Human Factors in Computing Systems},
    articleno = {590},
    numpages = {15},
    location = {New Orleans, LA, USA},
    series = {CHI '22}
}

@inproceedings{kaur2022recruit,
author = {Harjot Kaur and Sabrina Klivan and Daniel Votipka and Yasemin Acar and Sascha Fahl},
title = {Where to Recruit for Security Development Studies: Comparing Six Software Developer Samples},
booktitle = {31st USENIX Security Symposium (USENIX Security 22)},
year = {2022},
isbn = {978-1-939133-31-1},
address = {Boston, MA},
pages = {4041--4058},
url = {https://www.usenix.org/conference/usenixsecurity22/presentation/kaur},
publisher = {USENIX Association},
month = aug
}

@inproceedings{krause2023pushed,
author = {Alexander Krause and Jan H. Klemmer and Nicolas Huaman and Dominik Wermke and Yasemin Acar and Sascha Fahl},
title = {Pushed by Accident: A {Mixed-Methods} Study on Strategies of Handling Secret Information in Source Code Repositories},
booktitle = {32nd USENIX Security Symposium (USENIX Security 23)},
year = {2023},
isbn = {978-1-939133-37-3},
address = {Anaheim, CA},
pages = {2527--2544},
url = {https://www.usenix.org/conference/usenixsecurity23/presentation/krause},
publisher = {USENIX Association},
month = aug
}

@inproceedings{lipp2022empirical,
    author = {Lipp, Stephan and Banescu, Sebastian and Pretschner, Alexander},
    title = {An empirical study on the effectiveness of static C code analyzers for vulnerability detection},
    year = {2022},
    isbn = {9781450393799},
    publisher = {Association for Computing Machinery},
    address = {New York, NY, USA},
    url = {https://doi.org/10.1145/3533767.3534380},
    doi = {10.1145/3533767.3534380},
    booktitle = {Proceedings of the 31st ACM SIGSOFT International Symposium on Software Testing and Analysis},
    pages = {544–555},
    numpages = {12},
    location = {Virtual, South Korea},
    series = {ISSTA 2022}
}

@inproceedings{bai2019effective,
author = {Jia-Ju Bai and Julia Lawall and Qiu-Liang Chen and Shi-Min Hu},
title = {Effective Static Analysis of Concurrency {Use-After-Free} Bugs in Linux Device Drivers},
booktitle = {2019 USENIX Annual Technical Conference (USENIX ATC 19)},
year = {2019},
isbn = {978-1-939133-03-8},
address = {Renton, WA},
pages = {255-268},
url = {https://www.usenix.org/conference/atc19/presentation/bai},
publisher = {USENIX Association},
month = jul
}

@article{dossche2024context,
  title={A Context-Sensitive, Outlier-Based Static Analysis to Find Kernel Race Conditions},
  author={Dossche, Niels and Abrath, Bert and Coppens, Bart},
  journal={arXiv preprint arXiv:2404.00350},
  year={2024}
}

@article{sergeyuk2025using,
  title={Using AI-based coding assistants in practice: State of affairs, perceptions, and ways forward},
  author={Sergeyuk, Agnia and Golubev, Yaroslav and Bryksin, Timofey and Ahmed, Iftekhar},
  journal={Information and Software Technology},
  volume={178},
  pages={107610},
  year={2025},
  publisher={Elsevier}
}

@inproceedings{ebiwonjumi2026generative,
  title={Do Generative AI Tools Change How Developers Comment and Document Code?},
  author={Ebiwonjumi, Oluwatito Ebiwonjumi and Ahakonye, Love Allen Chijioke and Kim, Dong-Seong},
  booktitle={2026 International Conference on Artificial Intelligence in Information and Communication (ICAIIC)},
  pages={668--673},
  year={2026},
  organization={IEEE}
}

\end{document}